\documentclass[journal, twoside, web]{ieeecolor}
\usepackage{generic}
\usepackage[compress]{cite}
\usepackage{amsmath,amssymb,amsfonts}
\usepackage{algorithmic}
\usepackage{graphicx}
\usepackage{textcomp}
\usepackage{booktabs}
\usepackage{multirow}
\usepackage{color}
\usepackage{subfigure}
\usepackage{makecell}
\usepackage{inputenc}
\usepackage[switch]{lineno}

\newcommand{\ie}{\textit{i}.\textit{e}., }
\newcommand{\eg}{\textit{e}.\textit{g}., }
\usepackage{hyperref}
\hypersetup{
colorlinks = true,
linkcolor = red,
anchorcolor = red,
citecolor = red,
filecolor = red,
urlcolor = red,
pdfauthor=author}
\def\BibTeX{{\rm B\kern-.05em{\sc i\kern-.025em b}\kern-.08em
    T\kern-.1667em\lower.7ex\hbox{E}\kern-.125emX}}
\begin{document}


\title{An Arbitrary Scale Super-Resolution Approach for 3D MR Images via Implicit Neural Representation}

\author{Qing Wu, Yuwei Li, Yawen Sun, Yan Zhou, Hongjiang Wei, Jingyi Yu, \IEEEmembership{Fellow, IEEE},\\ and Yuyao Zhang, \IEEEmembership{Member, IEEE}
\thanks{This work was supported by National Natural Science Foundation of China (No. 62071299, 61901256, 91949120, and 81901693) and Shanghai ``Rising Stars of Medical Talent' Youth Development Program, Youth Medical Talents-Medical Imaging Practitioner Program (No. SHWRS (2020)\_087).}
\thanks{Qing Wu, Yuwei Li, and Jingyi Yu are with School of Information Science and Technology, ShanghaiTech University, Shanghai, China (e-mail: \{wuqing, liyw, yujingyi\}@shanghaitech.edu.cn).}
\thanks{Yawen Sun and Yan Zhou are with Department of Radiology, Ren Ji Hospital, School of Medicine, Shanghai Jiao Tong University, Shanghai 200127, P.R. China (e-mail: \{cjs1119, clare1475\}@hotmail.com).}
\thanks{Hongjiang Wei is with School of Biomedical Engineering and Institute of Medical Robotics, Shanghai Jiao Tong University, Shanghai, China (e-mail: hongjiang.wei@sjtu.edu.cn).}
\thanks{Yuyao Zhang (Corresponding Author) is with School of Information Science and Technology and iHuman Institute, Shanghaitech University, Shanghai, China (e-mail: zhangyy8@shanghaitech.edu.cn).}}

\maketitle
\begin{abstract}
\par High Resolution (HR) medical images provide rich anatomical structure details to facilitate early and accurate diagnosis. In magnetic resonance imaging (MRI), restricted by hardware capacity, scan time, and patient cooperation ability, isotropic 3-dimensional (3D) HR image acquisition typically requests long scan time and, results in small spatial coverage and low signal-to-noise ratio (SNR). Recent studies showed that, with deep convolutional neural networks, isotropic HR MR images could be recovered from low-resolution (LR) input via single image super-resolution (SISR) algorithms. However, most existing SISR methods tend to approach scale-specific projection between LR and HR images, thus these methods can only deal with fixed up-sampling rates. 
In this paper, we propose ArSSR, an \textit{\textbf{Ar}bitrary \textbf{S}cale \textbf{S}uper-\textbf{R}esolution} approach for recovering 3D HR MR images. 
In the ArSSR model, the LR image and the HR image are represented using the same implicit neural voxel function with different sampling rates. 
Due to the continuity of the learned implicit function, a \textit{single} ArSSR model is able to achieve arbitrary and infinite up-sampling rate reconstructions of HR images from any input LR image. 
Then the SR task is converted to approach the implicit voxel function via deep neural networks from a set of paired HR and LR training examples. The ArSSR model consists of an encoder network and a decoder network. Specifically, the convolutional encoder network is to extract feature maps from the LR input images and the fully-connected decoder network is to approximate the implicit voxel function. 
Experimental results on three datasets show that the ArSSR model can achieve state-of-the-art SR performance for 3D HR MR image reconstruction while using a single trained model to achieve arbitrary up-sampling scales. Code and data for this work are available at: \url{https://github.com/iwuqing/ArSSR}
\end{abstract}


\section{Introduction}
\label{sec:introduction}
\par Magnetic Resonance Imaging (MRI), as a non-invasive and radiation-free imaging technology, is commonly used for disease detection, diagnosis, and treatment monitoring. High-quality 3-dimensional (3D) high-resolution (HR) MR images provide rich tissue anatomical details, facilitating early-accurate diagnosis and quantitative image analysis. However, limited by the trade-off among image resolution, signal-to-noise ratio (SNR), and scanning time, it is difficult to acquire high-quality 3D isotropic HR MR images. For example, to increase the image resolution from $2\times 2\times 2$ mm$^3$ to $1\times 1\times 1$ mm$^3$ requires the average of 64 scans for maintaining a similar SNR, which thus significantly increases scanning time \cite{jia2017new, scherrer2012super}. To alleviate this trade-off, a common and effective strategy is to scan high-SNR low-resolution (LR) MR images and then improve the image resolution via super-resolution (SR) algorithms. A bunch of efficient and powerful SR algorithms have been proposed for this purpose. According to the number of demanded LR MR images \cite{wang2020deep}, SR algorithms can be briefly categorized into multi-image super-resolution (MISR) and single-image super-resolution (SISR). In MISR methods \cite{peled2001superresolution,jia2017new,ebner2020automated,scherrer2012super,IREM}, the HR image is generated by combining the high-frequency image information from multiple non-isotropic (\ie image with high in-plane resolution and thick slices) LR MR images, which are scanned from the same object in different scanning orientations. While SISR methods \cite{shi2015lrtv, rueda2013single} directly reconstruct the HR image from a single isotropic LR input image without additional scanning. In this paper, we focus our attention on the SISR method category.
\par Conventional SISR methods include interpolation-based methods and regularization-based methods. Interpolation-based methods, such as cubic/sinc interpolation, often suffer from blocking artifacts due to their poor de-aliasing ability \cite{jia2017new}; while regularization-based methods only achieve limited SR performance due to their over-idealistic prior assumptions \cite{chen2020mri}. Recently, the accuracy of the SISR task has been significantly improved by deep convolutional neural networks (CNNs) due to their non-linearity to learn the mapping between LR images and HR images. A good deal of CNN-based SISR methods \cite{SRCNN, DCSRN, ResCNN, chen2020mri, delannoy2020segsrgan, EAGAN, lyu2020multi, EGGAN} with well-designed network architectures for 3D MR images have been proposed. However, to the best of our knowledge, most of the existing deep-learning-based methods are designed to deal with the SR tasks of one or several fixed integer up-sampling scales (\eg $2\times$, $3\times$, and $4\times$). This is mainly because most models tend to approach a scale-specific projection between LR and HR images using deep neural networks. Therefore, these methods have to be respectively trained and stored for each up-sampling scale, which is very computation-expensive and resource-intensive. 
\par
\begin{figure}[t]
	\centering
	\includegraphics[width=0.465\textwidth]{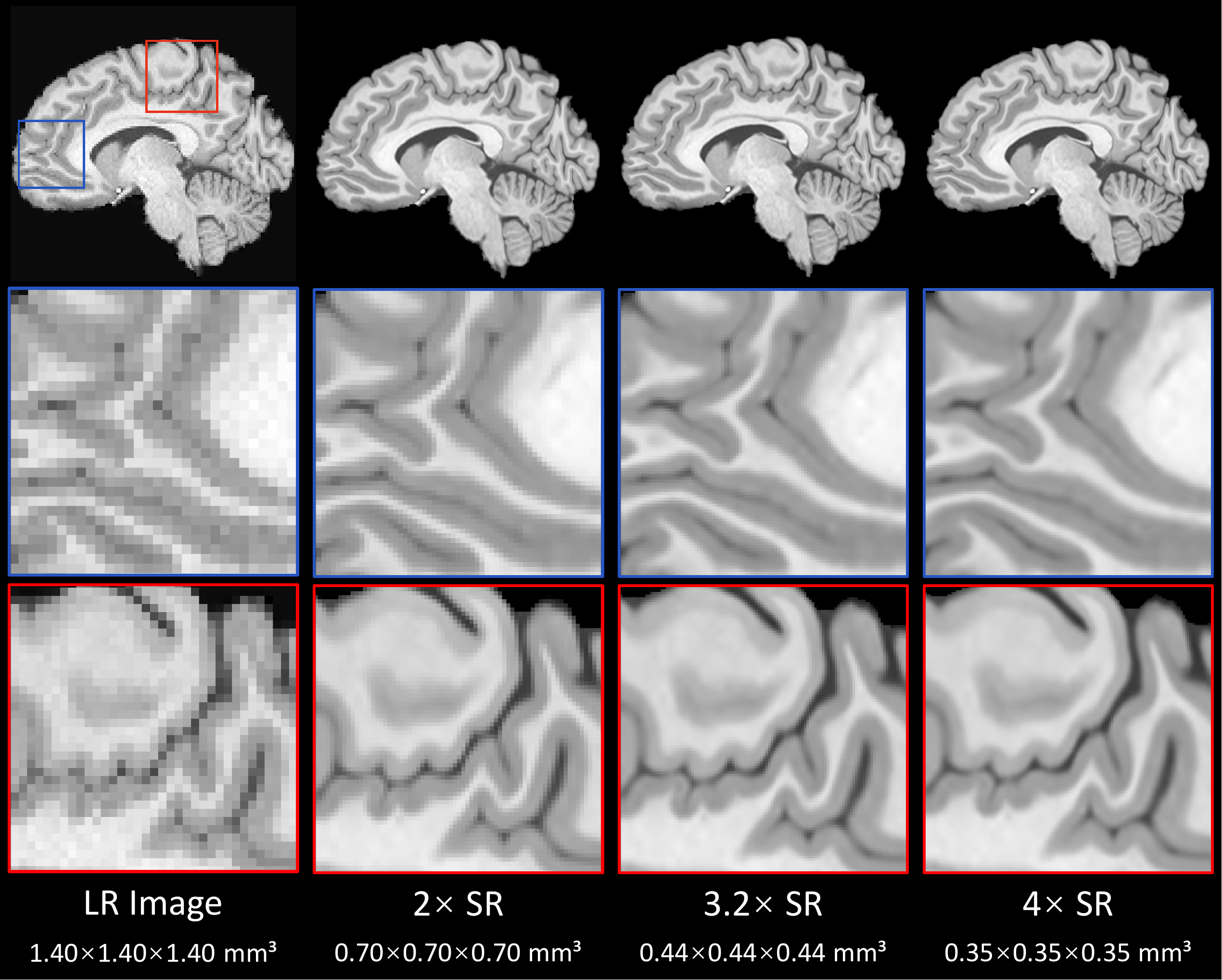}
	\caption{An example of the SISR tasks of three different isotropic up-sampling scales $k=\lbrace2,\ 3.2,\ 4\rbrace$ for a 3D brain MR image by the \textit{single} ArSSR model. Note that for the isotropic up-sample scale of $k$, the voxel size reduces by $k^{3}$ times, \ie $2\times$ SR denotes a voxel size  of 8 times smaller than before. }
	\label{figure-example}
\end{figure}

 In this paper, we propose ArSSR, an \textit{\textbf{Ar}bitrary \textbf{S}cale \textbf{S}uper-\textbf{R}esolution} model based on implicit neural representation for achieving high-quality 3D HR MR images. 
Inspired by a recent coordinated based subject-specific implicit neural network SR method IREM \cite{IREM}, which achieved superior SR performance and arbitrary up-scaling rates.
In the ArSSR model, we propose to learn a continuous implicit neural voxel function to represent images with different resolutions (LR and HR images) with different sampling rates, and extend the implicit function to overcome the subject-specific constraint of IREM.
Then, both LR and HR images are reconstructed as discrete sampling with different sampling rates of the same implicit voxel function.
Due to the continuity nature, a single implicit voxel function is able to produce arbitrary resolution images and thus lead the SR task to achieve arbitrary up-sampling scale rates. 
Then the SR task is thus converted to approach the implicit voxel function via deep neural networks from a set of paired HR and LR training examples. The ArSSR model consists of an encoder network and a decoder network. Specifically, the convolutional encoder network is to extract voxel-specific feature maps from the LR input images and the fully-connected decoder network is to approximate the implicit voxel function.
Figure \ref{figure-example} shows an example of the SISR tasks of different up-sampling rates for a 3D brain MR image by the proposed ArSSR model. It is able to conduct a variety of up-sampling rates including any integer rate or non-integer rate. Note that for the isotropic up-sample scale of $k$, the voxel size reduces by $k^{3}$ times, \ie $4\times$ SR denotes a voxel size of 64 times smaller than before. To comprehensively evaluate the SR performance, we conduct comparison experiments on three different datasets (including two simulation datasets and one real data collection dataset). Both qualitative and quantitative results demonstrate that the proposed ArSSR model can achieve state-of-the-art SR performance for 3D HR MR image reconstruction while using a single trained model to achieve arbitrary up-sampling scales.
\section{Related Work}
\label{section-related-work}

\subsection{Single Image Super-Resolution}

\par As a low-level task in computer vision, SISR has received widespread attention for decades \cite{wang2020deep}. SISR aims to reconstruct the corresponding HR image from a single LR input image. Conventional SISR methods include interpolation-based methods and regularization-based methods. Interpolation-based methods, such as cubic/sinc interpolation, are widely used due to their computational simplicity, but they often introduce severe blocking artifacts \cite{siu2012review, jia2017new}. Regularization-based methods \cite{shi2015lrtv, rudin1992nonlinear} recover the desired HR image by solving a convex optimization problem. Since the SISR task is a severely ill-posed problem (\ie there exist multiple HR solutions for a single LR input image), regularization-based methods usually adopt over-idealistic yet sophisticated prior assumptions to restrict the possible solution space. However, those prior assumptions do not always hold, which may greatly limit the SR performance of the models. For example, total variation \cite{rudin1992nonlinear} supposes that the optimal HR image should be constant in a small neighborhood, which is inconsistent with the fact in some regions (\eg brain cerebellum) of the HR images that have abundant local details.
\par Recently, deep learning methods have shown great potential in the SISR task. \cite{dong2015image} proposed a super-resolution convolutional neural network (SRCNN), which is the first deep-learning-based SISR model. SRCNN directly learns the optimal mapping between LR images and HR images from a large number of training examples using an end-to-end way thus it achieves  great performance improvement compared with the traditional interpolation- and regularization-based SISR models. After that, many deep-learning-based SISR models \cite{lim2017enhanced, kim2016accurate, SRGAN, shi2016real, RDN, ma2020structure, kim2016deeply, lai2017deep, mao2016image, du2020brain} with more complicated network structures have been proposed. For instance, \cite{kim2016accurate} developed a very deep convolutional network (VDSR) by increasing the network depth and utilizing residual connections. \cite{SRGAN} proposed a super-resolution generative adversarial network (SRGAN). Due to the perceptual loss function consisting of an adversarial loss and a content loss, SRGAN is able to reconstruct photo-realistic natural images for $4\times$ up-sampling scale.
\par Following the rapid development of deep learning networks in the SISR task of 2D natural images, many SR models have been adopted in the SISR task for 3D medical images. \cite{SRCNN} successfully reconstructed high-quality 3D HR MR brain images by SRCNN3D, a 3D version of the SRCNN \cite{dong2015image}. \cite{DCSRN} developed a densely connected super-resolution network (DCSRN) by incorporating dense connections. \cite{ResCNN} proposed a 20-layers residual convolutional neural network (ResCNN) by increasing the network depth and using residual connections. \cite{EAGAN} developed mDCSRN-GAN, the first generative adversarial SR network for 3D MR images. Benefiting from the well-designed network structures, all these models achieved satisfying SR performance. Yet, due to the fact that these models regard the SISR of different up-sampling scales as independent tasks, they have to be trained and stored for each up-sampling scale, respectively. Different from previous works, the proposed ArSSR model considers the SISR with arbitrary up-sampling scales as a single task. Therefore, the single fine-trained ArSSR model can conduct the SISR tasks of arbitrary scales, which greatly improves the practical applicability.
\begin{figure*}[t]
	\center
	\includegraphics[width=0.9\linewidth]{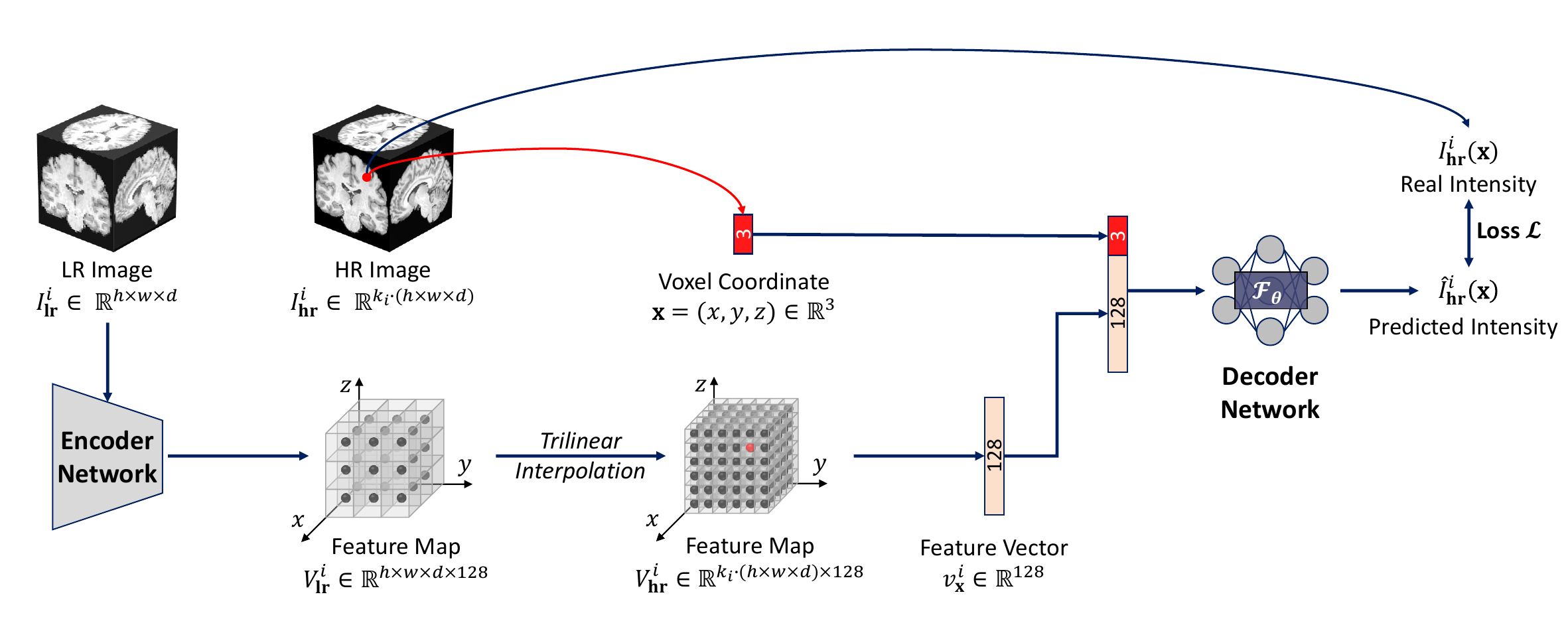}
	\caption{An overview of the proposed ArSSR model.}
	\label{method-pipeline}
\end{figure*}
\subsection{Implicit Neural Representation}
\par In the real world, most of the visual signals can be considered as continuous functions of a certain dimensional variable. For instance, a 2D natural image can be modeled by a continuous function of 2D pixel coordinates; a 3D MR/CT image can be modeled by a continuous function of 3D voxel coordinates. However, conventional signal representations are usually discrete (\eg pixel-based representations for 2D images and voxel-based representations for 3D MR/CT images) due to the digital computational system, which results in a severe trade-off between the accuracy of the represented signals and the storage resources. To alleviate this problem, implicit neural representation (INR) recently has emerged \cite{ImNet, DeepSDF, OccuNEt} as a novel way for continuously parametrizing a variety of signals. In the INR, a represented signal is modeled by a pre-defined continuous function that maps spatial coordinates to the signal responses. Then, the function is approximated by a simple fully connected network (\ie an MLP). Once the function is approached well, the MLP theoretically can estimate the signal response at any given position. Therefore, the INR is continuous and independent of the signal resolution.
\par The INR initially was proposed in these three works \cite{ImNet, DeepSDF, OccuNEt} for 3D shape and surface modeling. \cite{ImNet} proposed an implicit field decoder that greatly improves the visual quality of generative shape modeling by concatenating point coordinates with shape features, feeding both as input to the implicit decoder; \cite{DeepSDF} developed a continuous deep signed distance function that can enable the representation of complex shapes with small memory without discretization errors; \cite{OccuNEt} proposed an occupancy network, in which a 3D surface is modeled as the continuous decision boundary of a deep neural network classifier and thus the 3D surface at infinite resolution can be accurately encoded with limited memory footprint. The three works firstly demonstrated that the INR greatly outperforms grid-, point-cloud-, and mesh-based representation for 3D surface and shape modeling. Since then, a number of INR-based models \cite{ConvOccuNEt, NeRf, yu2021pixelnerf, liu2020neural, park2020deformable, peng2021neural} for 3D shape and surface have been proposed. In the field of 2D natural images, \cite{LIFF} proposed a local implicit image function trained by a self-supervised SR task that is theoretically able to recover HR images of arbitrary resolutions from a single LR input image; \cite{lin2021infinitygan} developed InfinityGAN model for infinite-resolution image synthesis; \cite{tang2021joint} proposed a novel joint implicit image function by formulating the guided SR task as a neural implicit image interpolation problem. Inspired by these works, we propose ArSSR, an INR-based arbitrary scale SISR model for 3D MR images. The single ArSSR model can conduct SISR tasks of arbitrary up-sampling scales by learning a novel implicit voxel function.

%
%
\section{Methodology}
\label{section-methodology}
\label{section:Methodology}
\subsection{Problem Modeling}

\par In our previous work \cite{IREM}, we suppose that the desired HR image $I_\mathbf{hr}$ and the observed LR image $I_\mathbf{lr}$ can be modeled via the same coordinated-based implicit function as below:
\begin{equation}
	I = f_\theta(\mathbf{x}) 
\end{equation}
where $\mathbf{x}=(x,y,z)$ denotes any 3D voxel spatial coordinate, and $I$ denotes the voxel intensity at the coordinate $\mathbf{x}$ in the image. Note that our coordinate system is slightly different from the physical-world coordinate system. Specifically, the physical-world coordinates are built on the spatial information of MR images (origin, direction, and spacing). Instead, our voxel coordinates are constructed based on the size of the MR images in the subject frame of reference and are normalized to the range of [-1, 1] along each dimension. The pair of HR and LR images are considered as the explicit representation of the implicit function $f_\theta(\mathbf{x})$ with different sampling rates. In the HR image $I_\mathbf{hr}$, more voxels are included than that in the LR image $I_\mathbf{lr}$. As we suppose the function $f_\theta(\mathbf{x})$ is continuous with the spatial coordinate system $\mathbf{x}$, once the function is well approached, an arbitrary up-sampling scale for the LR image can be achieved. 
\par However, it is impractical to train a specific function $f_\theta(\mathbf{x})$ for each image. Instead, we are eager to find a more flexible implicit function that is able to represent each pair of LR and HR images in the SR task. Towards this end, we propose a novel implicit voxel function $\mathcal{F}_\theta$ as below:
\begin{equation}
	I^i = \mathcal{F}_\theta(\mathbf{x}, v_\mathbf{x}^i)
	\label{e-function}
\end{equation}
where $v_\mathbf{x}^i \in \mathbb{R}^{128}$ denotes a voxel-specific latent space feature vector for the voxel intensity at the spatial coordinate $\mathbf{x}$ in image $I^i$. The vector is extracted using an SR task-specific convolutional encoder. Instead of simply representing an image via a spatial system, the variables of $\mathcal{F}_\theta$ include both coordinate and intensity embedding features of the MR images to be reconstructed. By combining convolutional encoders with implicit function decoders, our model is able to integrate the local semantic information in latent space, thus substantially improving the ability of the implicit decoder to represent and recover fine image details when scaling to any finer image resolution.
\subsection{Model Overview}

\par An overview of the proposed ArSSR model is illustrated in Figure \ref{method-pipeline}. For a given pair of LR-HR images from the training set $\mathbf{I} = \lbrace I_{\mathbf{lr}}^i \in \mathbb{R}^{h\times w\times d}, I_{\mathbf{hr}}^i \in \mathbb{R}^{k_i\cdot \left(h\times w\times d\right)}\rbrace_{i=1}^{M}$, where $M$ is the number of training examples and $k_i$ is the up-sampling scale for the $i$-th pair of LR-HR images, we first use a convolutional encoder network to convert LR image $I_\mathbf{lr}^i$ into a feature map $V_\mathbf{lr}^i\in \mathbb{R}^{h\times w\times d\times 128}$, where each element is a feature vector $v_\mathbf{x}^i\in \mathbb{R}^{128}$ for the corresponding voxel at the coordinate $\mathbf{x}$ in the LR image $I_\mathbf{lr}^i$. Then, for any query of 3D voxel coordinate $\mathbf{x}=(x, y, z)$ from the HR image $I_\mathbf{hr}^i$, we generate its feature map $V_\mathbf{hr}^i\in \mathbb{R}^{k_i\cdot \left(h\times w\times d\right)\times 128}$ by trilinear interpolating on the LR image feature map $V_\mathbf{lr}^i$. Finally, the decoder network takes the query coordinate $\mathbf{x}$ and the corresponding feature vector $v_\mathbf{x}^i$ in the feature map $V_\mathbf{hr}^i$ as input, and outputs the voxel intensity $\hat{I}_\mathbf{hr}^i(\mathbf{x})$ at the spatial coordinate $\mathbf{x}$. By using back-propagating gradient decent algorithm \cite{rumelhart1986learning} to minimize the L1 loss function $\mathcal{L}$ between the predicted voxel intensity $\hat{I}_\mathbf{hr}^i(\mathbf{x})$ and the real voxel intensity $I_\mathbf{hr}^i(\mathbf{x})$, the encoder and decoder network can be optimized simultaneously. Formally, the L1 loss function $\mathcal{L}$ is expressed as follows:
\begin{equation}
	\label{loss}
	\mathcal{L} = \frac{1}{N* \mathcal{K}}\sum_{i=1}^{N}\sum_{j=1}^{\mathcal{K}}{\mid I_\mathbf{hr}^i(\mathbf{x}_{j})-\hat{I}_\mathbf{hr}^i(\mathbf{x}_{j})\mid}
\end{equation}
where $N$ denotes the number of the HR-LR image pairs and $\mathcal{K}$ denotes the number of the coordinates sampled from each HR image $I_\mathbf{hr}^i$ during each training iteration.
\par After the model training, the encoder network acts as a feature map extractor and the decoder network is an approximator of the implicit voxel function $\mathcal{F}_\theta$. For any unknown LR image $I_\mathbf{lr} \in\mathbb{R}^{h\times w\times d}$, we firstly use the encoder network to extract its feature map $V_\mathbf{lr}\in \mathbb{R}^{h\times w\times d\times 128}$, and then we build a more dense coordinate grid for generating a HR image $I_\mathbf{hr} \in\mathbb{R}^{H\times W\times D}$, where $[H,W,D] =k\cdot[h,w,d]$ and $k>1$ is an arbitrarily up-sampling scale, with higher sampling rate on the implicit image function. The feature map $V_\mathbf{hr}\in \mathbb{R}^{H\times W\times D\times 128}$ for the HR image is synthesized via trilinear interpolating on the LR image feature map $V_\mathbf{lr}$. And afterward, we can use the decoder network to predict voxel intensity $I_\mathbf{hr}(\mathbf{x})$ at any spatial coordinate in the reconstructed HR image grid. Thus the HR image $I_\mathbf{hr}$ with an arbitrary spatial resolution can be achieved.
\subsection{Encoder Network}

\par In pioneering works \cite{ImNet, DeepSDF, OccuNEt} that represent 3D surfaces using implicit neural representation, a general and efficient strategy is to encode input signals (\eg volume data, point cloud) into a single latent vector by an encoder network. However, this strategy does not integrate local information in the input signals as claimed in \cite{ConvOccuNEt, LIFF}, so these models cannot finely represent the 3D surfaces with complex shapes. To this end, \cite{ConvOccuNEt} developed a convolutional occupancy network that conducts elements-wise feature extraction for the input signals, which greatly improves the performance for 3D surface reconstruction. Meanwhile, \cite{LIFF} proposed a local implicit image function for representing 2D natural images, which can recover HR images with an arbitrary resolution from an unknown LR image.
\par Inspired by the works \cite{ConvOccuNEt, LIFF}, we use an encoder network to extract voxel-wise latent space feature vectors for the LR images. The encoder network takes the LR image $I_\mathbf{lr}^i\in \mathbb{R}^{h\times w \times d}$ as input and outputs a feature map $V_\mathbf{lr}^i\in \mathbb{R}^{h\times w\times d\times 128}$, where each element is a feature vector $v_\mathbf{x}^i \in \mathbb{R}^{128}$ for the corresponding voxel at the coordinate $\mathbf{x}$ in the LR image $I_\mathbf{lr}^i$. This voxel-wise latent space extraction strategy facilitates the following decoder network to efficiently integrate local image intensity information and thus successfully recover fine image details in the HR image at even a large up-sampling scale. We apply a modified residual dense network (RDN) \cite{RDN} to implement our encoder network. In particular, we remove the upscaling layer in the original RDN, extend all the 2D convolutional layers to 3D convolutional layers, and set the output channel number of the last layer as 128. The number of residual dense blocks (RDBs) is 8 and each RDB has 3 3D convolutional layers. The growth rate of RDBs is set to 64. Note that our encoder network can be replaced by any other SR network without upscaling layer. In the experiment section, the effect of the encoder network is discussed.
\begin{figure}[t]
	\center
	\includegraphics[width=0.37\textwidth]{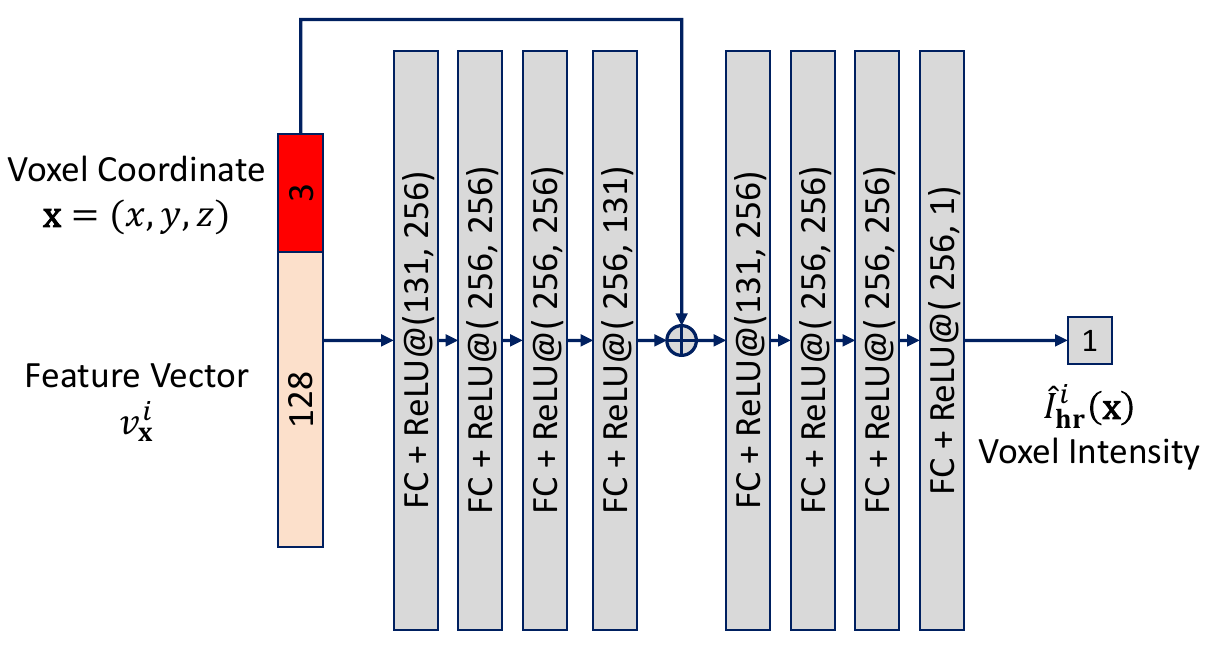}
	\caption{Architecture of the decoder network. The decoder network takes a query voxel coordinate $\mathbf{x}=(x,y,z)$ and the corresponding feature vectors $v_{\mathbf{x}}^i$ as input, and outputs the voxel intensity $\hat{I}_\mathbf{hr}^i(\mathbf{x})$ at that spatial coordinate $\mathbf{x}$.}
	\label{method-decoder}
\end{figure}
\subsection{Decoder Network}
\par After the encoder network, the LR image $I_\mathbf{lr}^i$ is converted to the feature map $V^i_\mathbf{lr}$. Our decoder network aims to approximate the implicit voxel function $\mathcal{F}_\theta$ and thus is able to estimate the HR image voxel intensities $\hat{I}_\mathbf{hr}^i(\mathbf{x})$ at any spatial coordinate $\mathbf{x}$. Specifically, as the HR image involves much more voxels than that in the input LR image, we need to fill the gaps for the enlarging number of feature vectors $v_\mathbf{x}^i$. Thus given any query of 3D voxel coordinate $\mathbf{x}$ from the HR image $I_\mathbf{hr}^i$, We generate its feature vector $v_{\mathbf{x}}^i$ by trilinear interpolating the feature map $V^i_\mathbf{lr}$. Formally, the feature vector $v_{\mathbf{x}}^i$ is calculated as:
\begin{equation}
    v_{\mathbf{x}}^i = \sum_{k=0}^{7}{\frac{S_{k}}{S}\cdot v_{k}^i}, \quad S = \sum_{k=0}^{7}{S_{k}}
    \label{equ-trilinear}
\end{equation}
where $v_k^i$ ($k=0,\cdots, 7$) are the nearest 8 feature vectors of the query coordinate $\mathbf{x}$ in the feature map $V^i_\mathbf{lr}$, $S_k$ is the volume of the cuboid between the query coordinate $\mathbf{x}$ and the coordinate diagonal to the feature vector $v_k^i$. Then, the decoder network takes each query voxel coordinate $\mathbf{x}$ and the feature vector $v_\mathbf{x}^i$ as input, and finally outputs the estimated voxel intensity $\hat{I}_\mathbf{hr}^i(\mathbf{x})$ at the coordinate $\mathbf{x}$. Figure \ref{method-decoder} illustrates the architecture of the decoder network. It consists of eight standard fully-connected (FC) layers, and a ReLU activation layer follows each FC layer. Two numbers in each gray box are the input channel and output channel of the FC layer. To facilitate the training process, we involve a residual connection between the input of the decoder network and the output of the 4th ReLU activation layer.
%
%
\section{Experiments}
\label{section-experiments}
\par To comprehensively evaluate the proposed ArSSR model, we conduct the following five experiments:

\subsubsection{The SISR Tasks on HCP-1200 Dataset} The ArSSR model and four baseline models (Cubic interpolation and three most-cited deep-learning-based MRI SISR models \cite{SRCNN, DCSRN, ResCNN}) are trained on training set and validation set of the Human Connectome Projection (HCP-1200) Dataset \cite{HCP-1200}. Then, on test set of the HCP-1200 dataset \cite{HCP-1200}, these well-trained models are used to conduct the SISR tasks of five isotropic up-sampling scales $k=\lbrace2,\ 2.5,\ 3,\ 3.5,\ 4\rbrace$. Specially, for an isotropic up-sampling scale of $k$, the voxel size reduces by $k^{3}$ times, which is very challenging task. For example, to perform a $4\times$ SR denotes to generate an image with voxel size of 64 times smaller than that before the SR process. Both the qualitative and quantitative results are reported. It is worth to notice that for the ArSSR, we train one model for achieving the five up-sampling scales; while for the other three deep-learning-based baselines, we train a model for each up-sampling scale, respectively.
\subsubsection{The SISR Tasks on Lesion Brain Dataset} To further compare the generalization ability of the ArSSR model with the four baseline methods, we employ all the well-trained comparison models to perform the SISR tasks of three isotropic up-sampling scales $k=\lbrace2,\ 3,\ 4\rbrace$ on the Lesion Brain dataset for assessing the ability to recover fine image detail that is not included in the training dataset. The qualitative and quantitative results are reported. 
\subsubsection{The SISR Tasks on Healthy Brain Dataset} To test the performance of the ArSSR and the baseline models in a real MR image scanning scenario, we use Cubic interpolation, ResCNN, and the ArSSR to conduct the SISR tasks of three isotropic up-sampling scales $k=\lbrace2,\ 3,\ 4\rbrace$ on the Healthy Brain dataset, which is a real collected dataset. We qualitatively demonstrate the difference between the SISR results and reference images. 
\subsubsection{Fully Automatic Segmentation based Evaluation} To further compare the quality of the SR results from all the different deep learning models, we follow the segmentation evaluation strategy as in the previous works \cite{EGGAN, chen2020mri}. Specifically, we use Advanced Normalization Tools (ANTs) (\url{http://stnava.github.io/ANTs/}), an open-source medical image processing tool, to conduct a fully automatic segmentation on the GT images and all the SISR results. Each brain MR image is separated into six regions. We quantitatively calculate Dice Similarity Coefficient (Dice) between the segmentation results on the GT and SR images. 
\subsubsection{Influence of Encoder Network} On the HCP-1200 dataset \cite{HCP-1200}, we investigate the effect of the encoder network on the performance of the proposed ArSSR model. Specifically, we employ three different CNNs (ResCNN \cite{ResCNN}, SRResNet \cite{SRGAN}, and RDN \cite{RDN}) to implement the encoder network. We qualitatively and quantitatively report the performance of the ArSSR models with the three encoder networks for the SISR tasks of the five scales.
\subsection{Experimental Settings}

\subsubsection{Datasets \& Preprocessing}
\par All experiments are conducted on three datasets: HCP-1200 dataset \cite{HCP-1200}, Lesion Brain dataset, and Healthy Brain dataset. The HCP-1200 \cite{HCP-1200} and Lesion Brain datasets are simulation datasets, where the LR images were generated by down-sampling from the corresponding HR images. In contrast, the Healthy Brain dataset is a real-world dataset, where LR and HR images were collected from the same subject using different image scanning protocols on a 3T MR scanner. The HCP-1200 dataset \cite{HCP-1200} we downloaded is skull-stripped, while the Lesion Brain and Healthy Brain datasets are not skull-stripped. We directly train and test our model on the datasets without any further processing on the skull.
\par \textit{\textbf{HCP-1200 Dataset}}: The HCP-1200 dataset \cite{HCP-1200} is a large public human brain MR image dataset, which consists of 1113 3D isotropic HR T1-weighted brain MR images. These HR T1-weighted MR images were acquired on Siemens 3T platforms using a 32-channel head coil. MPRAGE sequences were used with the following parameters: repetition time (TR) $=2400$ ms; echo time (TE) $=2.14$ ms; field of view (FOV) $=224\times224$ mm$^2$; matrix size $=320\times320$; slice thickness $=0.7$ mm; $256$ continuous sagittal slices; voxel size $=0.7\times0.7\times0.7$ mm$^3$. The HCP-1200 dataset is split into three parts: $780$ images for training set, $111$ ones in validation set, and $222$ ones in test set. Note that the test set is only used for final performance evaluation and no images are overlapped in the three parts. Additionally, we apply cropping and padding to all the original images to modify the image dimension size from $320\times320\times256$ to $264\times264\times264$ for the convenience of quantitative evaluations. For the test set, we first conduct data normalization on the intensity of the raw MR images to build ground truth (GT) HR images and then downsample the GT images by cubic interpolation at the specific down-sampling scales $k=\lbrace2,\ 2.5,\ 3,\ 3.5,\ 4\rbrace$ to generate the LR images. Thus the test set is prepared. For the training and validation set, we follow the pipeline of the data preprocessing in the previous work \cite{LIFF}: (\romannumeral1) We first conduct data normalization on the raw images of $264^3$ size; (\romannumeral2) We then apply random cropping to extract 6 image patches of a $40^3$ dimension size; (\romannumeral3) Thirdly, we conduct center cropping to generate HR patches of a $(10\times k)^3$ dimension size; (\romannumeral4) Finally, we downsample the HR patches by cubic interpolation to simulate the LR patches of a $10^3$ dimension size. For the proposed ArSSR model, the scale $k$ is randomly sampled from the uniform distribution $\mathcal{U}(2, 4)$.
\par \textit{\textbf{Lesion Brain Dataset}}: Seventeen T1-weighted brain MR images are acquired from 17 patients with cerebrovascular disease (CVD) on a 3T GE scanner with an 8-channel head coil. To be more specific, the CVD subjects in the lesion brain dataset are cerebral small vessel disease patients caused by arteriolosclerosis. MPRAGE sequences were used with following parameters: TR $=5.5$ ms; TE $=1.7$ ms; flip angle $=15^{\circ}$; matrix size $=256\times256$; slice thickness $=1$ mm; 156 continuous sagittal slices; voxel size $=1\times1\times1$ mm$^3$. The current study was approved by the Research Ethics Committee of Ren Ji Hospital and School of Medicine, Shanghai Jiao Tong University, China. The Lesion Brain dataset is only used as a test dataset for evaluating the ability of the proposed method on recovering HR images with unknown image detail (\ie brain tissue lesion pattern). All included images comprise of brain tissue lesions in white matter (WM), which have never been seen during the model training procedure. We use the original images as GT images, and then downsample the GT images by cubic interpolation with the specific down-sampling scales $k=\lbrace2,\ 3,\ 4\rbrace$ to simulate the LR images.
\begin{figure}[t]
    \centering
    \includegraphics[width=0.45\textwidth]{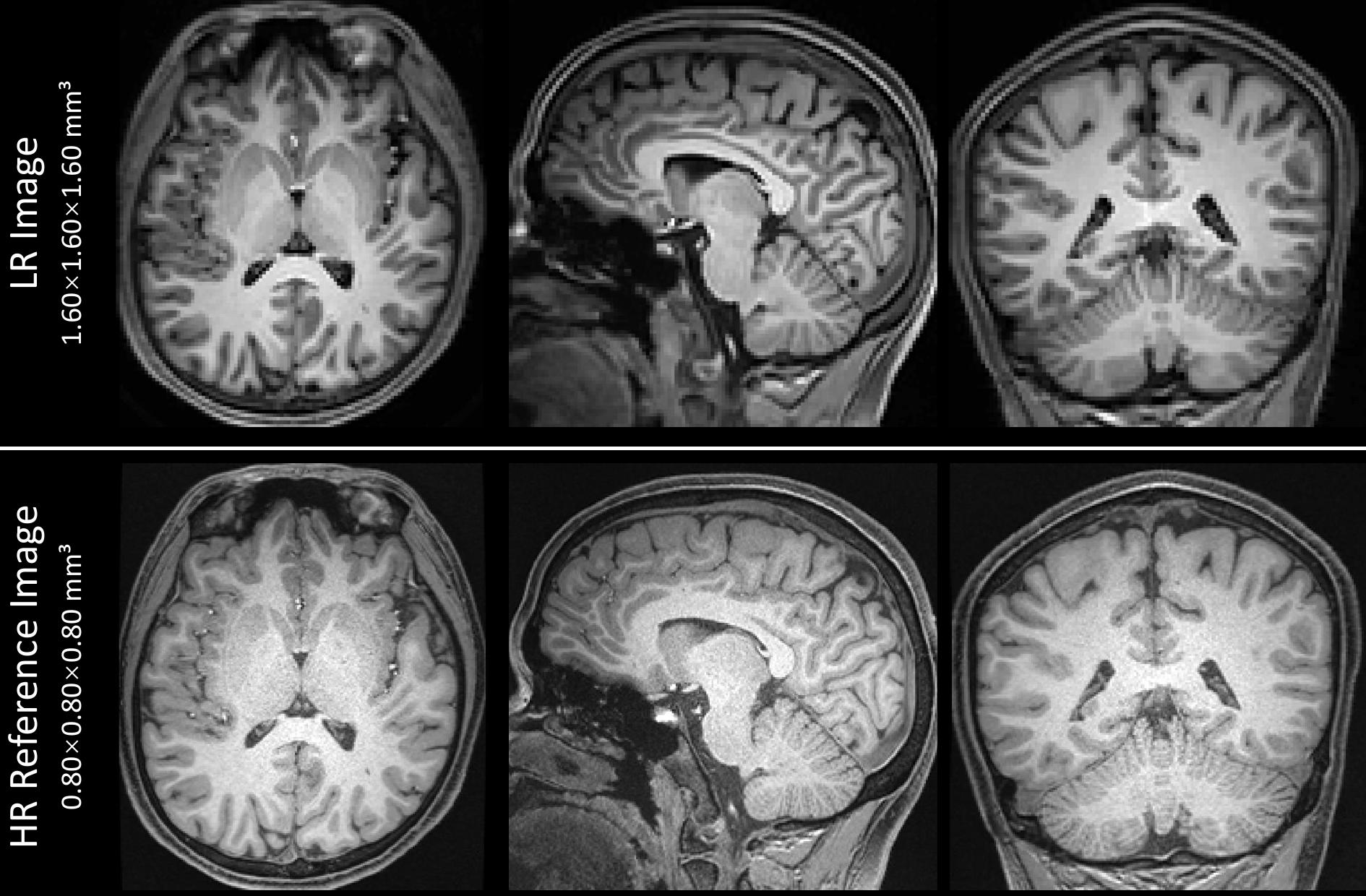}
    \caption{Qualitative comparison of LR image and HR reference image from the Healthy Brain dataset. LR image has a spatial resolution of $1.6\times1.6\times1.6$ mm$^3$, scanning time $\approx3$ minutes; reference image has a spatial resolution of $0.8\times0.8\times0.8$ mm$^3$, scanning time $\approx11$ minutes.}
    \label{real_data}
\end{figure}
\par \textit{\textbf{Healthy Brain Dataset}}: We scan two T1-weighted brain MR images from the same healthy adult volunteer on a 3T United Imaging MR scanner. As shown in Figure \ref{real_data}, the first row is a LR image with high SNR, while the second row is a HR image with low SNR. The LR image was acquired by MPRAGE sequence with the following parameters: TR $=6.70$ ms; TE $=3.0$ ms; flip angle $=8^{\circ}$; matrix size $=160\times150$; slice thickness $=1.6$ mm; 104 continuous sagittal slices; voxel size $=1.6\times1.6\times1.6$ mm$^3$; scanning time $\approx3$ minutes. The HR image was acquired by sequence with the following parameters: TR $=8.07$ ms; TE $=3.4$ ms; flip angle $=8^{\circ}$; matrix size $=320\times300$; slice thickness $=0.8$ mm; $208$ continuous sagittal slices; voxel size $=0.8\times0.8\times0.8$ mm$^3$; scanning time $\approx11$ minutes. This data collection was approved by the Ethics Committee of ShanghaiTech University. In our experiments, we use the LR image as the input image for the models and the HR image as the reference image.
\begin{figure*}[t]
	\centering
	\includegraphics[width=0.85\textwidth]{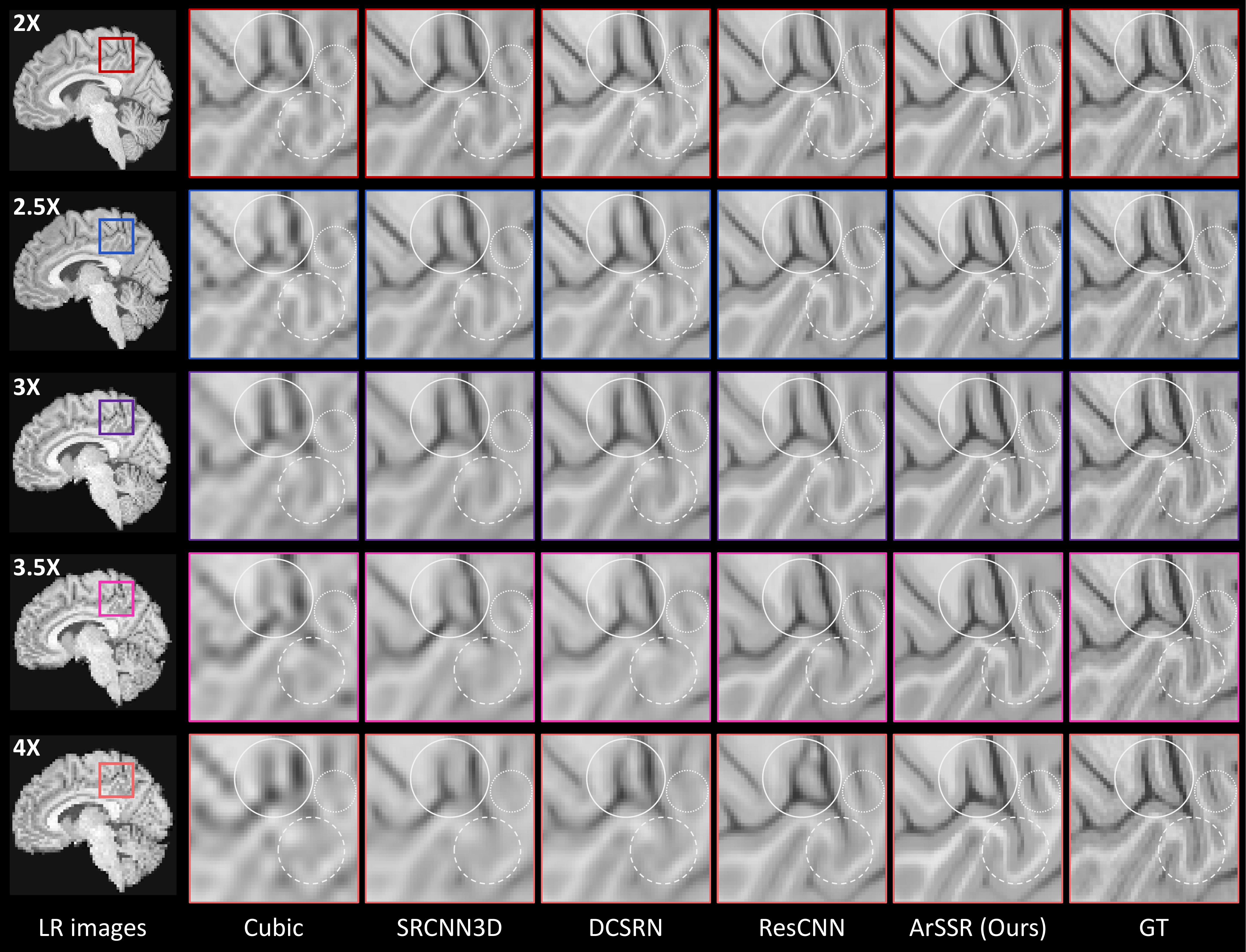}
	\caption{Qualitative results of all the compared models on a test sample (No. \#827052) from the test set of the HCP-1200 dataset \cite{HCP-1200} for the SISR tasks of the five up-sampling scales $k=\lbrace2,\ 2.5,\ 3,\ 3.5,\ 4\rbrace$.}
	\label{figure-ex1}
\end{figure*}
\subsubsection{Evaluation Metrics}
\par To quantitatively evaluate the performance of all the models for the SISR task comparison, we adopt five objective image quality metrics. We first compute Peak Signal-to-Noise Ratio (PSNR) and Similarity Index Measure (SSIM) \cite{SSIM}. They are widely adopted for the quantitative evaluation of the SR models. However, the recent works \cite{EAGAN, SRGAN, EGGAN, LPIPS} proved that the two metrics cannot represent image visual quality well (\eg an over-smoothed SISR result with an extreme loss on the fine image details may score high on the two metrics). Therefore, we also calculate Learned Perceptual Image Patch Similarity (LPIPS \cite{LPIPS}), a deep-learning-based perceptual similarity measure for quantitative evaluation. By comprehensive experiments, \cite{LPIPS} proved that compared with pixel-wise metrics like PSNR and NMSE, the LPIPS represents image visual quality more realistically. Additionally, two non-referenced quality metrics (Perceptual Sharpness Index (PSI \cite{PSI}) and Local Phase Coherence-based Sharpness Index (LPC-SI \cite{LPC-SI})) are also calculated. LPIPS \cite{LPIPS}, PSI \cite{PSI}, and LPC-SI \cite{LPC-SI} are exclusively designed for 2D images, while the MR images discussed in this paper are in 3D. Thus we use the slice-by-slice strategy to compute them. Specifically, all three metrics are calculated in three steps: Given a 3D MR image, we first extract its 2D MR slices from three orthogonal directions (Axial, Sagittal, and Coronal directions). Then, we compute the scores for each 2D MR slice. Finally, we average the scores of all the 2D MR slices to calculate the final scores.
\subsubsection{Training Details}
\par For the training of the ArSSR model, in each step of training, we first randomly sample 15 LR-HR patch pairs (\ie $N=15$ in Equ. \ref{loss}) from the training set, and then we randomly sample 8000 voxel coordinates (\ie $\mathcal{K}=8000$ in Equ. \ref{loss}) from each HR patch as the inputs of the decoder network. We use Adam optimizer \cite{Adam} to minimize the L1 loss function and the hyperparameters of the Adam are set as follows: $\beta_1=0.9,\ \beta_2=0.999,\ \varepsilon={10}^{-8}$. The learning rate starts from ${10}^{-4}$ and decays by a factor of 0.5 every 200 epochs. The total number of training epochs is 2500, which takes about 12 hours on a single NVIDIA TITAN RTX 24G GPU. The best model is saved by checkpoints during the training.
\subsubsection{Compared Methods}
\par We adopt cubic interpolation and three popular deep-learning-based SISR models (SRCNN3D \cite{SRCNN}, DCSRN \cite{DCSRN}, and ResCNN \cite{ResCNN})  for 3D MR images as compared methods. Cubic interpolation is implemented by the scipy library \cite{2020SciPy-NMeth} of Python, while the three deep-learning-based models are implemented by Pytorch deep learning framework \cite{paszke2019pytorch} following the original papers. For the three models, we employ Adam optimizer \cite{Adam} to train them through the back-propagation algorithm with a mini-batch size of 10, and the hyper-parameters of the Adam optimizer \cite{Adam} are set as follows: $\beta_{1}=0.9$, $\beta_{2}=0.999$, and $ \varepsilon=10^{-8}$. The leaning rate is initialized as $10^{-4}$ and decays by factor 0.5 every 200 epochs. The total number of training epochs is 2000 and the best models are saved by checkpoints during the training.
\subsection{The SISR Tasks on the HCP-1200 Dataset}
\begin{figure}[t]
	\centering
	\includegraphics[width=0.35\textwidth]{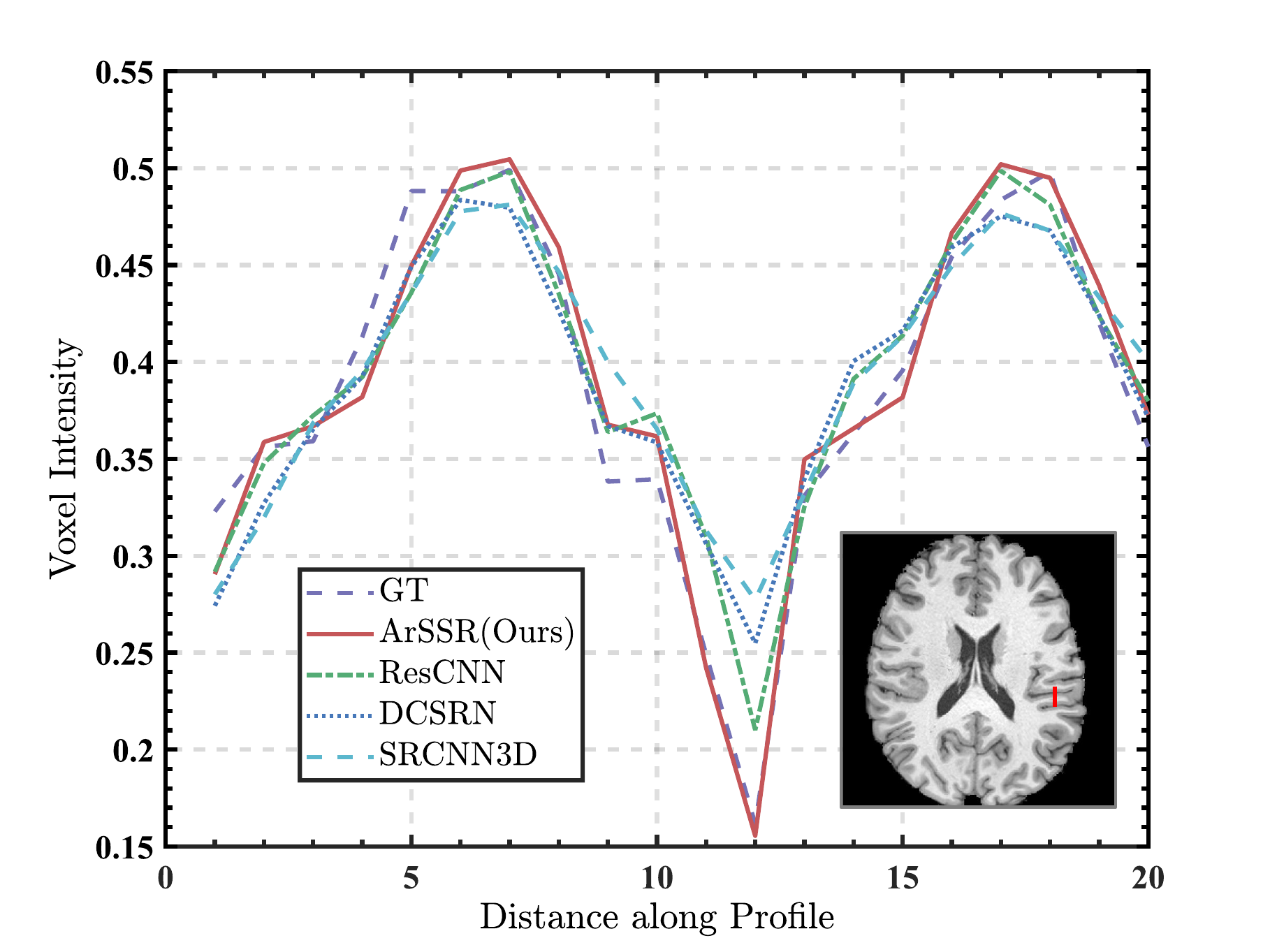}
	\caption{Intensity profiles of the selected red lines on GT image and the $3\times$ SR results on a test sample (No. \#531536) from the test set of the HCP-1200 dataset \cite{HCP-1200}.}
	\label{figure-ex1-3x}
\end{figure}
\begin{table}[t]
\caption{Quantitative results of all the compared models on the HCP-1200 dataset \cite{HCP-1200} for the SISR tasks of the five up-sampling scales $k=\lbrace2,\ 2.5,\ 3,\ 3.5,\ 4\rbrace$.}
\label{table-ex1}
\centering
\scalebox{0.8}{
\begin{tabular}{c|c|c|c|c|c|c} 
\toprule
\textbf{Scales} & \textbf{Models}& \textbf{PSNR$\uparrow$}& \textbf{SSIM$\uparrow$}& \textbf{LPIPS$\downarrow$}& \textbf{PSI$\uparrow$}& \textbf{LCP-SI$\uparrow$} \\
\hline
\multirow{5}{*}{$2\times$}             
& Cubic&                            
$24.44$& 
$0.9553$& 
$0.0954$& 
$0.2978$& 
$0.9359$\\ 
& SRCNN3D&                           
$38.11$& 
$0.9764$& 
$0.0634$& 
$0.2959$& 
$0.9372$\\
& DCSRN&                             
\textcolor{blue}{$38.50$}& 
\textcolor{blue}{$0.9783$}&  
\textcolor{blue}{$0.0540$}& 
\textcolor{blue}{$0.3125$}&
$0.9400$\\ 
& ResCNN&                            
\textcolor{red}{$39.58$}& 
\textcolor{red}{$0.9824$}& 
\textcolor{red}{$0.0371$}& 
\textcolor{red}{$0.3355$}&
\textcolor{red}{$0.9474$}\\ 
& ArSSR (Ours)&                                
$36.55$&      
$0.9700$&        
$0.0549$&      
$0.3083$&
\textcolor{blue}{$0.9435$}\\
\hline\hline 
\multirow{5}{*}{$2.5\times$}           
& Cubic&                            
$23.28$&      
$0.9389$&      
$0.1565$&      
$0.2560$&
$0.9289$\\ 
& SRCNN3D&                          
\textcolor{blue}{$36.01$}&      
\textcolor{blue}{$0.9613$}&      
$0.1001$&      
$0.2666$&
$0.9301$\\
& DCSRN&                            
$35.96$&      
$0.9287$&      
$0.0992$&      
$0.2722$&
$0.9360$\\ 
& ResCNN&                           
\textcolor{red}{$37.20$}&      
\textcolor{red}{$0.9665$}&      
\textcolor{red}{$0.0621$}&      
\textcolor{red}{$0.3315$}&
\textcolor{red}{$0.9465$}\\ 
& ArSSR (Ours)&                           
$34.57$&      
$0.9538$&      
\textcolor{blue}{$0.0715$}&      
\textcolor{blue}{$0.3089$}&
\textcolor{blue}{$0.9428$}\\ 
\hline\hline 
\multirow{5}{*}{$3\times$}             
& Cubic&                            
$22.54$&      
$0.9268$&      
$0.2196$&      
$0.2182$&
$0.9224$\\ 
& SRCNN3D&                          
$35.00$&      
$0.9506$&      
$0.1303$&      
$0.2269$&
$0.9243$\\ 
& DCSRN&                            
\textcolor{blue}{$35.41$}&      
\textcolor{blue}{$0.9523$}&      
$0.1168$&     
$0.2507$&
$0.9314$\\ 
& ResCNN&                           
\textcolor{red}{$35.98$}&      
\textcolor{red}{$0.9622$}&      
\textcolor{blue}{$0.0876$}&      
\textcolor{blue}{$0.2741$}&
\textcolor{blue}{$0.9408$}\\ 
& ArSSR (Ours)&                           
$33.42$&      
$0.9426$&      
\textcolor{red}{$0.0811$}&      
\textcolor{red}{$0.2986$}&
\textcolor{red}{$0.9426$}\\ 

\hline\hline 
\multirow{5}{*}{$3.5\times$}           
& Cubic&                            
$21.95$&      
$0.9120$&      
$0.2765$&      
$0.1938$&
$0.9115$\\ 
& SRCNN3D&                          
\textcolor{blue}{$33.76$}&      
\textcolor{blue}{$0.9341$}&      
$0.1643$&      
$0.1904$&
$0.9156$\\ 
& DCSRN&                            
$33.60$&      
$0.7558$&      
$0.1537$&      
$0.2108$&
$0.9263$\\ 
& ResCNN&                           
\textcolor{red}{$34.30$}&      
\textcolor{red}{$0.9466$}&      
\textcolor{blue}{$0.1160$}&      
\textcolor{blue}{$0.2707$}&
\textcolor{blue}{$0.9396$}\\ 
& ArSSR (Ours)&                            
$31.98$&     
$0.9242$&      
\textcolor{red}{$0.1009$}&      
\textcolor{red}{$0.2798$}&
\textcolor{red}{$0.9405$}\\ 

\hline\hline 
\multirow{5}{*}{$4\times$}             
& Cubic&                            
$21.33$&     
$0.9037$&      
$0.3252$&      
$0.1729$&
$0.9062$\\ 
& SRCNN3D&                          
$33.24$&      
\textcolor{blue}{$0.9261$}&      
$0.1881$&      
$0.1675$&
$0.9116$\\ 
& DCSRN&                            
\textcolor{blue}{$33.52$}&      
$0.9175$&      
$0.1714$&      
$0.2007$&
$0.9242$\\ 
& ResCNN&                           
\textcolor{red}{$33.86$}&      
\textcolor{red}{$0.9383$}&      
\textcolor{blue}{$0.1389$}&      
\textcolor{blue}{$0.2349$}&
\textcolor{blue}{$0.9342$}\\ 
& ArSSR (Ours)&                             
$31.33$&      
$0.9155$&      
\textcolor{red}{$0.1136$}&      
\textcolor{red}{$0.2534$}&
\textcolor{red}{$0.9394$}\\
\bottomrule
\end{tabular}}
\end{table}
\begin{figure*}[t]
	\centering
	\includegraphics[width=0.85\textwidth]{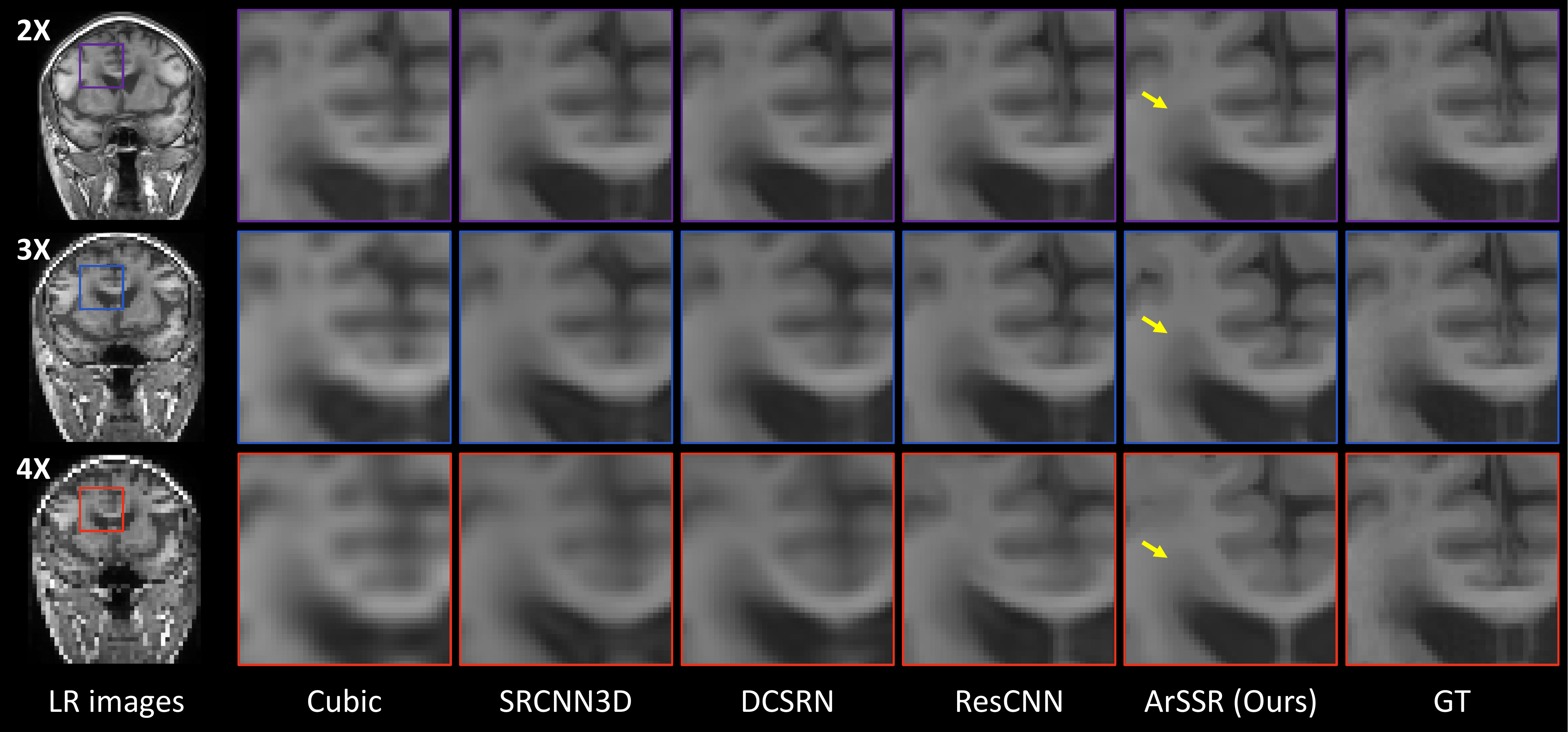}
	\caption{Qualitative results of all the compared models on a test sample (No. \#3) from the lesion brain dataset for the SISR tasks of the three up-sampling scales $k=\lbrace2,\ 3,\ 4\rbrace$.}
	\label{figure-ex-lesion}
\end{figure*}
\par Figure \ref{figure-ex1} shows the qualitative results. To sum up, the ArSSR model produces the best SISR results in terms of image sharpness and HR fine details for all the scales. Specifically, the SISR results from cubic interpolation are extremely blurry and with severe artifacts; while SRCNN3D \cite{SRCNN}, DCSRN \cite{DCSRN}, and ResCNN \cite{ResCNN} models recover limited fine details in HR image and their results get more blurred with the up-sample scale increasing. For example, as emphasized in the round dash mark, the narrow CSF between-in the cortical gyrus is only well recovered by the proposed model in each scaling rate. As shown in Figure \ref{figure-ex1-3x}, we select a red line on the axial view of the $3\times$ SISR results (No. \#531536) from all the comparison models to evaluate the reconstructed local image contrast. The SR result by the proposed ArSSR model has the sharpest changes in the intensity profile line, which is the closest result compared with that in the GT image. In Table \ref{table-ex1}, we demonstrate the quantitative results. In terms of PSNR and SSIM \cite{SSIM}, ResCNN \cite{ResCNN} achieves the best performance for all the scale cases, while our ArSSR model only scores higher than cubic interpolation. However, we need to emphasize that, as claimed in \cite{EAGAN, EGGAN, SRGAN}, an over-smoothed SR result may score high on the two metrics. Therefore, the reason why SRCNN3D \cite{SRCNN}, DCSRN \cite{DCSRN}, and ResCNN \cite{ResCNN} achieve higher scores than the proposed ArSSR may be that they produced more smooth SISR results. For LPIPS \cite{LPIPS}, LPC-SI \cite{LPC-SI}, and PSI \cite{PSI}, we observe that the ArSSR provides the best performance in most of the scale cases. To be more specific, the improvement from the ArSSR model is more significant with the up-sample scale increasing, which is consistent with the above qualitative results.
\begin{table}[t]
\caption{Quantitative results of all the compared models on the lesion brain dataset for the SISR tasks of the three up-sampling scales $k=\lbrace2,\ 3,\ 4\rbrace$.}
\label{table-ex1-lesion}
\centering
\scalebox{0.8}{
\begin{tabular}{c|c|c|c|c|c|c} 
\toprule
\textbf{Scales} & \textbf{Models}& \textbf{PSNR$\uparrow$}& \textbf{SSIM$\uparrow$}& \textbf{LPIPS$\downarrow$}& \textbf{PSI$\uparrow$}& \textbf{LCP-SI$\uparrow$} \\
\hline
\multirow{3}{*}{$2\times$}             
& Cubic&                            
$33.00$& 
$0.8362$& 
$0.0471$&
$0.2402$&
$0.9650$\\ 
& SRCNN3D&                           
$34.56$& 
$0.9669$& 
$0.0296$& 
$0.2289$&
$0.9674$\\
& DCSRN&                             
\textcolor{blue}{$35.53$}& 
\textcolor{red}{$0.9777$}&  
\textcolor{blue}{$0.0187$}& 
$0.2317$&
\textcolor{blue}{$0.9693$}\\ 
& ResCNN&                            
\textcolor{red}{$35.70$}& 
\textcolor{blue}{$0.9759$}& 
\textcolor{red}{$0.0149$}& 
\textcolor{red}{$0.2439$}&
\textcolor{red}{$0.9703$}\\ 
& ArSSR (Ours)&                                
$31.27$&      
$0.9582$&        
$0.0245$&      
\textcolor{blue}{$0.2373$}&
$0.9683$\\
\hline\hline 
\multirow{3}{*}{$3\times$}             
& Cubic&                            
$28.36$&      
$0.8001$&      
$0.1739$&      
$0.2038$&
$0.9554$\\ 
& SRCNN3D&                          
\textcolor{red}{$29.26$}&      
$0.8206$&      
$0.1006$&      
$0.1898$&
$0.9556$\\ 
& DCSRN&                            
\textcolor{blue}{$30.01$}&      
\textcolor{red}{$0.9284$}&      
$0.0780$&     
$0.2042$&
$0.9600$\\ 
& ResCNN&                           
$29.13$&      
\textcolor{blue}{$0.9125$}&      
\textcolor{blue}{$0.0617$}&      
\textcolor{blue}{$0.2408$}&
\textcolor{blue}{$0.9678$}\\ 
& ArSSR (Ours)&                           
$27.10$&      
$0.8975$&      
\textcolor{red}{$0.0603$}&      
\textcolor{red}{$0.2507$}&
\textcolor{red}{$0.9678$}\\ 
\hline\hline 
\multirow{3}{*}{$4\times$}             
& Cubic&                            
$25.82$&     
$0.7508$&      
$0.2798$&      
$0.1711$&
$0.9430$\\ 
& SRCNN3D&                          
\textcolor{blue}{$26.35$}&      
$0.8027$&      
$0.1767$&      
$0.1725$&
$0.9403$\\ 
& DCSRN&                            
\textcolor{red}{$26.99$}&      
\textcolor{red}{$0.8588$}&      
$0.1580$&      
$0.1869$&
$0.9519$\\ 
& ResCNN&                           
$25.95$&      
\textcolor{blue}{$0.8416$}&      
\textcolor{blue}{$0.1394$}&      
\textcolor{red}{$0.2274$}&
\textcolor{blue}{$0.9593$}\\ 
& ArSSR (Ours)&                             
$25.16$&      
$0.8388$&      
\textcolor{red}{$0.1005$}&      
\textcolor{blue}{$0.2145$}&
\textcolor{red}{$0.9666$}\\
\bottomrule
\end{tabular}}
\end{table}
\begin{figure*}[t]
    \centering
    \includegraphics[width=\textwidth]{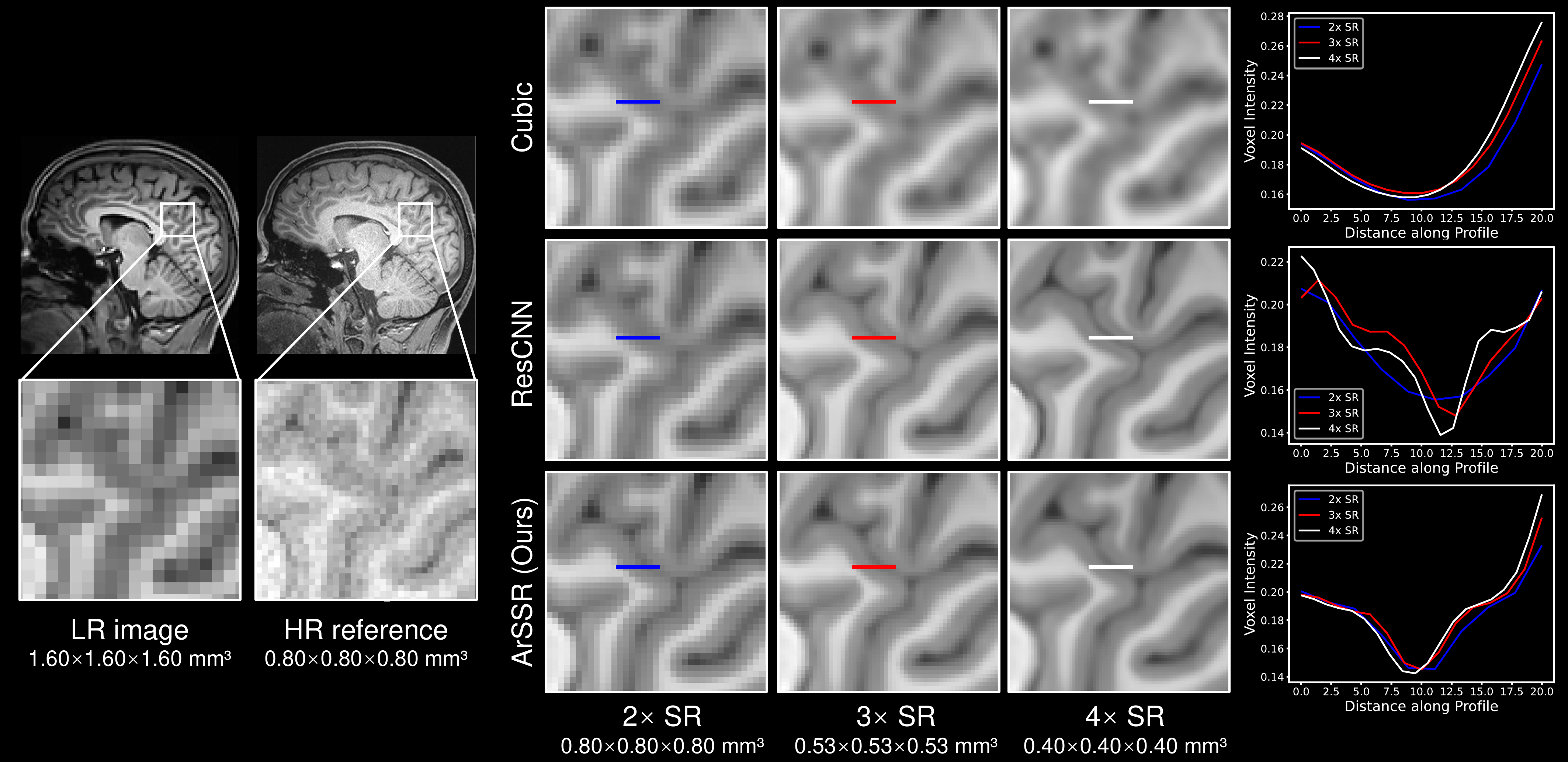}
    \caption{Qualitative results of LR input image, Reference image, SISR results by cubic interpolation, ResCNN \cite{ResCNN}, and ArSSR (Ours) on the Healthy Brain dataset for the SISR tasks of the three up-sampling scales $k=\{2,\ 3,\ 4\}$.}
    \label{figure-real}
\end{figure*}
\subsection{The SISR Tasks on the Lesion Brain Dataset}
\par Figure \ref{figure-ex-lesion} shows the qualitative results. Affected by different MR scanners and image scanning parameters, the image contrast between Gray Matter (GM) and White Matter (WM) in this dataset is slightly different from that in the training dataset (HCP-1200 Dataset \cite{HCP-1200}). Obviously, the SISR results from cubic interpolation are still with severe blocking artifacts. While the SISR results from the three deep-learning-based models dramatically become more blurry with the up-sampling scale increasing. As indicated by the yellow arrows in Figure \ref{figure-ex-lesion}, when the scale rate $k=\lbrace3,\ 4\rbrace$, the WM lesion beside cerebrospinal fluid (CSF) is tended to merge into the CSF region in SR results of ResCNN \cite{ResCNN} and SRCNN3D \cite{SRCNN}, and in DSCRN \cite{DCSRN} the boundary of lesion and CSF is also very hard to be figured out. Besides, these image intensities become brighter in the CSF and the borders between CSF and GM are also blurred in the three deep-learning-based models. Only ArSSR model can produce sharp and clear image contrast between GM and WM, and between CSF and GM in the cortical regions (shown in the triangle mark) for all three up-sampling scales. Table \ref{table-ex1-lesion} shows the quantitative results. Similar to the quantitative result on the test set of the HCP-1200 dataset \cite{HCP-1200}, in terms of PSNR and SSIM\cite{SSIM}, the proposed ArSSR model has low scores; while in terms of LPIPS \cite{LPIPS}, PSI \cite{PSI}, and LCP-SI \cite{LPC-SI}, the ArSSR model obtain the best performance in the most of the scales. Although when $k=\lbrace3,\ 4\rbrace$ the contrasts of brain lesion are not as clear as in the GT, but the boundary between lesion, normal WM tissue, and CSF are clearly distinguishable, while the contrasts in GM, WM, and CSF are very clear in the triangle bounding box.
\begin{table}[t]
    \renewcommand\arraystretch{1.15}
    \caption{Qualitative results (PSI/LCP-SI) of cubic interpolation, ResCNN \cite{ResCNN}, and ArSSR (Ours) on the Healthy Brain dataset for the SISR tasks of the three scales $k=\{2,\ 3,\ 4\}$.}
    \centering
    \scalebox{0.90}{
    \begin{tabular}{c|c|c|c}
        \toprule
        \textbf{Scales} & \textbf{Cubic}& \textbf{ResCNN}& \textbf{ArSSR (Ours)} \\
        \hline
         $2\times$&$0.2737/0.9430$&$0.2878/\textcolor{red}{0.9482}$&$\textcolor{red}{0.3071}/0.9438$\\
         $3\times$&$0.1953/0.9279$&$0.1990/\textcolor{red}{0.9338}$&$\textcolor{red}{0.2387}/0.9336$ \\
         $4\times$&$0.1452/0.9043$&$0.1600/0.9204$&$\textcolor{red}{0.1701/0.9284}$ \\\hline
         Mean&$0.2047/0.9251$&$0.2156/0.9341$&$\textcolor{red}{0.2386/0.9353}$ \\
         \bottomrule
    \end{tabular}}
    \label{tab:real}
\end{table}
\begin{figure*}[t]
	\centering
	\includegraphics[width=0.87\textwidth]{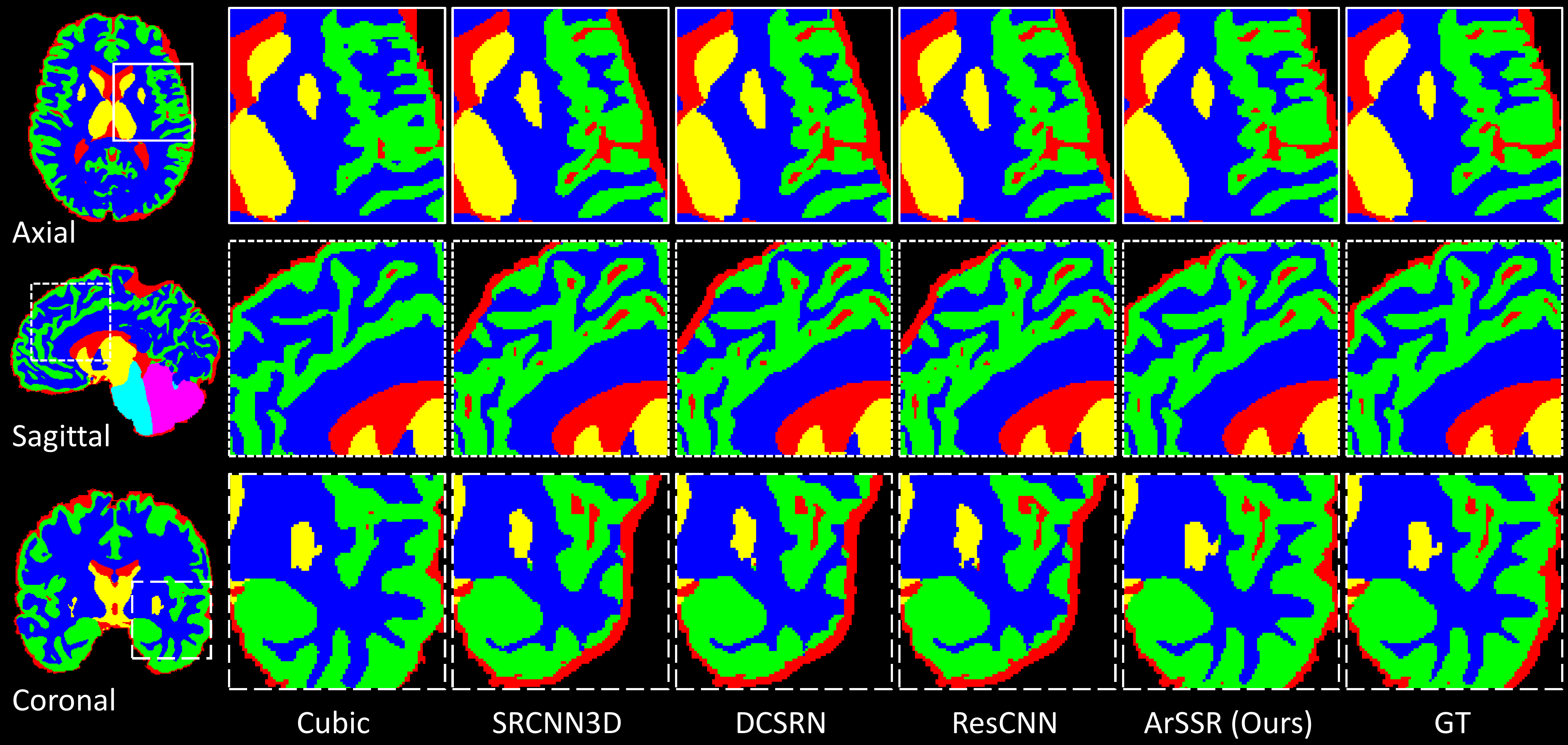}
	\caption{Qualitative results of the fully automatic segmentation on a test sample (No. \#996782) from the $2\times$ SISR results of all the compared models on the test dataset of the HCP-1200 dataset \cite{HCP-1200}. }
	\label{figure-ex2}
\end{figure*}
\begin{table}[t]
\caption{Quantitative results (Dice) of the fully automatic segmentation on the SISR results of all the compared models on the HCP-1200 dataset \cite{HCP-1200}.}
\label{table-ex2}
\centering
\scalebox{0.8}{
\begin{tabular}{c|c|c|c|c|c|c} 
\toprule
\textbf{Scales}                           & \textbf{Regions} & \textbf{Cubic}                  & \textbf{SRCNN3D}                & \textbf{DCSRN}                  & \textbf{ResCNN}                 & \textbf{ArSSR}            \\ 
\hline
\multirow{7}{*}{$2\times$}   & CSF                          & \textcolor{blue}{$0.6857$} & $0.6840$                   & $0.6806$                   & $0.6816$                   & \textcolor{red}{$0.9065$}   \\ 

                                          & GM                  & $0.7285$                   & $0.8639$                   & $0.8636$                   & \textcolor{blue}{$0.8650$} & \textcolor{red}{$0.9491$}   \\ 

                                          & WM                 & $0.8499$                   & $0.9510$                   & $0.9512$                   & \textcolor{blue}{$0.9517$} & \textcolor{red}{$0.9739$}   \\ 

                                          & DGM             & $0.8896$                   & $0.9499$                   & \textcolor{blue}{$0.9510$} & $0.9508$                   & \textcolor{red}{$0.9751$}   \\ 

                                          & Brain Stem                   & $0.9137$                   & \textcolor{blue}{$0.9561$} & $0.9560$                   & $0.9557$                   & \textcolor{red}{$0.9804$}   \\ 

                                          & Cerebellum                   & \textcolor{blue}{$0.8640$} & $0.8465$                   & $0.8456$                   & $0.8476$                   & \textcolor{red}{$0.9776$}   \\ 
\cline{2-7}
                                          & Means                   & $0.8219$                   & $0.8752$                   & $0.8747$                   & \textcolor{blue}{$0.8754$} & \textcolor{red}{$0.9604$}   \\ 
\hline\hline
\multirow{7}{*}{$2.5\times$} & CSF                          & \textcolor{blue}{$0.7362$} & $0.6731$                   & $0.6750$                   & $0.6746$                   & \textcolor{red}{$0.8797$}   \\ 

                                          & GM                  & $0.7854$                   & $0.8527$                   & $0.8554$                   & \textcolor{blue}{$0.8608$} & \textcolor{red}{$0.9294$}   \\ 

                                          & WM                 & $0.8796$                   & $0.9454$                   & $0.9475$                   & \textcolor{blue}{$0.9503$} & \textcolor{red}{$0.9626$}   \\ 

                                          & DGM             & $0.9070$                   & $0.9400$                   & $0.9440$                   & \textcolor{blue}{$0.9472$} & \textcolor{red}{$0.9620$}   \\ 

                                          & Brain Stem                   & $0.9378$                   & $0.9591$                   & \textcolor{blue}{$0.9603$} & $0.9595$                   & \textcolor{red}{$0.9806$}   \\ 

                                          & Cerebellum                   & \textcolor{blue}{$0.9066$} & $0.8476$                   & $0.8495$                   & $0.8502$                   & \textcolor{red}{$0.9723$}   \\ 
\cline{2-7}
                                          & Means                   & $0.8588$                   & $0.8696$                   & $0.8720$                   & \textcolor{blue}{$0.8738$} & \textcolor{red}{$0.9478$}   \\ 
\hline\hline
\multirow{7}{*}{$3\times$}   & CSF                          & \textcolor{blue}{$0.7009$} & $0.6663$                   & $0.6692$                   & $0.6714$                   & \textcolor{red}{$0.8570$}   \\ 

                                          & GM                  & $0.7603$                   & $0.8377$                   & $0.8480$                   & \textcolor{blue}{$0.8559$} & \textcolor{red}{$0.9118$}   \\ 

                                          & WM                 & $0.8665$                   & $0.9352$                   & $0.9414$                   & \textcolor{blue}{$0.9457$} & \textcolor{red}{$0.9526$}   \\ 

                                          & DGM             & $0.8926$                   & $0.9292$                   & $0.9350$                   & \textcolor{blue}{$0.9370$} & \textcolor{red}{$0.9519$}   \\ 

                                          & Brain Stem                   & $0.9270$                   & $0.9514$                   & $0.9522$                   & \textcolor{blue}{$0.9533$} & \textcolor{red}{$0.9711$}   \\ 

                                          & Cerebellum                   & \textcolor{blue}{$0.8883$} & $0.8401$                   & $0.8401$                   & $0.8434$                   & \textcolor{red}{$0.9653$}   \\ 
\cline{2-7}
                                          & Means                   & $0.8393$                   & $0.8600$                   & $0.8643$                   & \textcolor{blue}{$0.8678$} & \textcolor{red}{$0.9350$}   \\ 
\hline\hline
\multirow{7}{*}{$3.5\times$} & CSF                          & \textcolor{blue}{$0.7337$} & $0.6548$                   & $0.6605$                   & $0.6523$                   & \textcolor{red}{$0.8308$}   \\ 

                                          & GM                  & $0.7932$                   & $0.8178$                   & $0.8323$                   & \textcolor{blue}{$0.8422$} & \textcolor{red}{$0.8909$}   \\ 

                                          & WM                 & $0.8860$                   & $0.9236$                   & $0.9338$                   & \textcolor{blue}{$0.9397$} & \textcolor{red}{$0.9406$}   \\ 

                                          & DGM             & $0.8972$                   & $0.9115$                   & $0.9212$                   & \textcolor{blue}{$0.9303$} & \textcolor{red}{$0.9404$}   \\ 

                                          & Brain Stem                   & $0.9415$                   & $0.9537$                   & $0.9549$                   & \textcolor{blue}{$0.9551$} & \textcolor{red}{$0.9712$}   \\ 

                                          & Cerebellum                   & \textcolor{blue}{$0.9126$} & $0.8416$                   & $0.8430$                   & $0.8438$                   & \textcolor{red}{$0.9636$}   \\ 
\cline{2-7}
                                          & Means                   & \textcolor{blue}{$0.8607$} & $0.8505$                   & $0.8576$                   & $0.8606$ & \textcolor{red}{$0.9229$}   \\ 
\hline\hline
\multirow{7}{*}{$4\times$}   & CSF                          & \textcolor{blue}{$0.6957$} & $0.6461$                   & $0.6581$                   & $0.6516$                   & \textcolor{red}{$0.8062$}   \\ 

                                          & GM                  & $0.7601$                   & $0.7967$                   & $0.8247$                   & \textcolor{blue}{$0.8312$} & \textcolor{red}{$0.8709$}   \\ 

                                          & WM                 & $0.8675$                   & $0.9095$                   & $0.9242$                   & \textcolor{red}{$0.9317$}  & \textcolor{blue}{$0.9284$}  \\ 

                                          & DGM             & $0.8813$                   & $0.8962$                   & $0.9121$                   & \textcolor{blue}{$0.9217$} & \textcolor{red}{$0.9291$}   \\ 

                                          & Brain Stem                   & $0.9265$                   & $0.9436$                   & $0.9470$                   & \textcolor{blue}{$0.9497$} & \textcolor{red}{$0.9618$}   \\ 

                                          & Cerebellum                   & \textcolor{blue}{$0.8862$} & $0.8340$                   & $0.8378$                   & $0.8384$                   & \textcolor{red}{$0.9529$}   \\ 
\cline{2-7}
                                          & Means                   & $0.8362$                   & $0.8377$                   & $0.8506$                   & \textcolor{blue}{$0.8540$} & \textcolor{red}{$0.9082$}   \\
\bottomrule
\end{tabular}}
\end{table}
\subsection{The SISR Tasks on the Healthy Brain Dataset}
\par Figure \ref{figure-real} shows the qualitative results. From the visual inspection, the SR results from cubic interpolation are blurry and rarely improved compared to the LR input image. ResCNN \cite{ResCNN} and the ArSSR model produce similar SISR results in terms of both the image sharpness and HR fine details for all three scales. While for the reconstruction of fine image details, our ArSSR model generates more consistent local image patterns in different up-sampling scales. Benefiting from the large voxel size in LR image scanning, the SNR in each voxel of the LR input image is about 8 times higher than that in the reference HR image. Therefore, the image contrast between white matter and gray matter in the reconstructed images is much higher than that in the reference HR image. In addition, we also demonstrate the intensity profiles of the CSF regions in-between cortical sulci and gyrus in the SR images. As shown in Figure \ref{figure-real}, cubic interpolation produces over-blurry results, where the intensity profile curves are very flat in three cases, while ResCNN obtains diverse results at the three up-sampling scales. In particular, ResCNN produces a relatively smooth image at scale $k=2$, but the SR results at scales $k=\{2,\ 3\}$ have sharp intensity changes in the CSF region. Only the SR results of our ArSSR model at the three scales have sharp intensity changes. More importantly, these changes are significantly consistent at different scales, which indicates the superior performance of our ArSSR model for arbitrary-scale SR. For quantitative comparison, an HR GT image does not exist since the HR reference and LR image of the real Healthy Brain dataset are scanned from two independent acquisitions. Therefore, we additionally compute the qualitative results of three methods in terms of PSI \cite{PSI} and LCP-SI \cite{LPC-SI}, two non-reference-based metrics, as shown in Table \ref{tab:real}. We can see that our ArSSR model generally obtains the best performance at three up-sampling scales.
\subsection{Fully Automatic Segmentation based Evaluation}
\par Figure \ref{figure-ex2} shows the qualitative results. Note that Red, Blue, Yellow, Pink, Green and Cyan represent CSF, Gray Matter (GM), Deep Gray Matter (DGM), Cerebellum, White Matter (WM), and Brain Stem, respectively. All three views (Axial view, Sagittal view, and Coronal view) clearly illustrate that compared with the baseline methods, the segmentation results of the ArSSR model are closest to those of the GT image. Table \ref{table-ex2} shows the quantitative results. Except for that when the scale $k=4$, ResCNN \cite{ResCNN} slightly outperforms the ArSSR model on the segmentation region of the While Matter (Dice: $0.9317$ vs $0.9284$), our ArSSR model achieves the best performance on all segmentation regions at all up-sampling scales.
\begin{figure}[t]
	\centering
	\includegraphics[width=0.48\textwidth]{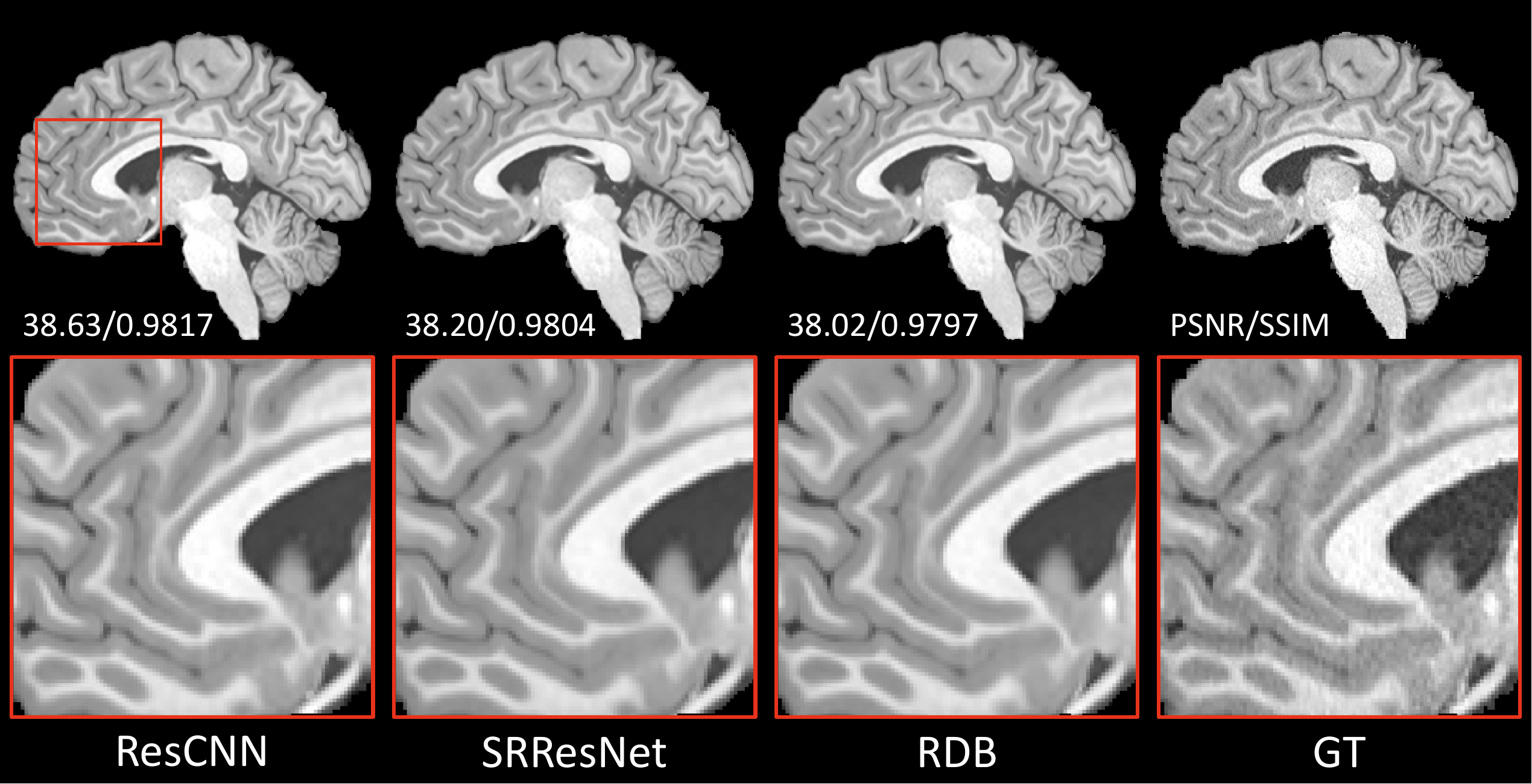}
	\caption{Qualitative results of the ArSSR models with three different encoder networks on a test sample (No. \#531536) from the test set of the HCP-1200 dataset \cite{HCP-1200} for the SISR task of the scale $k=2$.}
	\label{figure-ex3}
\end{figure}
\begin{table}[ht]
\caption{Quantitative results of the ArSSR models with three different encoder networks (RDN \cite{RDN}, ResCNN \cite{ResCNN}, SRResNet \cite{SRGAN}) on the test set of the HCP-1200 dataset \cite{HCP-1200} for the SISR tasks of the five up-sampling scales $k=\lbrace2,\ 2.5,\ 3,\ 3.5,\ 4\rbrace$.}
\label{table-ex3}
\centering
\scalebox{0.8}{
\begin{tabular}{c|c|c|c|c|c|c} 
\toprule
\textbf{Scales} & \textbf{Models}& \textbf{PSNR$\uparrow$}& \textbf{SSIM$\uparrow$}& \textbf{LPIPS$\downarrow$}& \textbf{PSI$\uparrow$}& \textbf{LCP-SI$\uparrow$} \\
\hline
\multirow{3}{*}{$2\times$}             
& ResCNN&                             
\textcolor{red}{${36.92}$}& 
\textcolor{red}{${0.9722}$}&  
$0.0573$& 
$0.3010$&
$0.9435$\\ 
& SRResNet&                            
$36.59$& 
$0.9708$& 
$0.0554$& 
$0.2937$&
\textcolor{red}{${0.9444}$}\\ 
& RDB&                             
$36.55$&      
$0.9700$&        
\textcolor{red}{${0.0549}$}&      
\textcolor{red}{${0.3083}$}&
$0.9435$\\

\hline\hline 
\multirow{3}{*}{$2.5\times$}           
& ResCNN&                             
\textcolor{red}{${35.17}$}& 
\textcolor{red}{${0.9585}$}&  
$0.0736$& 
$0.3043$&
$0.9429$\\ 
& SRResNet&                            
$34.85$& 
$0.9562$& 
$0.0730$& 
$0.2910$&
\textcolor{red}{${0.9435}$}\\ 
& RDB&                             
$34.57$&      
$0.9538$&      
\textcolor{red}{${0.0715}$}&      
\textcolor{red}{${0.3089}$}&
$0.9428$\\ 
\hline\hline 
\multirow{3}{*}{$3\times$}             
& ResCNN&                             
\textcolor{red}{${34.10}$}& 
\textcolor{red}{${0.9485}$}&  
$0.0841$& 
$0.2862$&
$0.9434$\\ 
& SRResNet&                            
$33.81$& 
$0.9462$& 
$0.0835$& 
$0.2753$&
\textcolor{red}{${0.9436}$}\\ 
& RDB&                             
$33.42$&      
$0.9426$&      
\textcolor{red}{${0.0811}$}&      
\textcolor{red}{${0.2986}$}&
$0.9426$\\ 
\hline\hline 
\multirow{3}{*}{$3.5\times$}           
& ResCNN&                             
\textcolor{red}{${32.77}$}& 
\textcolor{red}{${0.9322}$}&  
$0.1054$& 
$0.2632$&
$0.9420$\\ 
& SRResNet&                            
$32.45$& 
$0.9295$& 
$0.1034$& 
$0.2486$&
\textcolor{red}{${0.9427}$}\\ 
& RDB&                             
$31.98$&     
$0.9242$&      
\textcolor{red}{${0.1009}$}&      
\textcolor{red}{${0.2798}$}&
$0.9405$\\ 
\hline\hline
\multirow{3}{*}{$4\times$}             
& ResCNN&                             
\textcolor{red}{${32.18}$}& 
\textcolor{red}{${0.9240}$}&  
$0.1187$& 
$0.2345$&
$0.9404$\\ 
& SRResNet&                            
$31.86$& 
$0.9220$& 
$0.1161$& 
$0.2220$&
\textcolor{red}{${0.9411}$}\\ 
& RDB&                             
$31.33$&      
$0.9155$&      
\textcolor{red}{${0.1136}$}&      
\textcolor{red}{${0.2534}$}&
$0.9394$\\
\bottomrule
\end{tabular}}
\end{table}
\subsection{Influence of Encoder Network}
Figure \ref{figure-ex3} illustrates the qualitative results of the ArSSR models with the three different encoder networks. We observe that, compared with the GT image, all the SISR results are less noisy and almost have the same HR fine details, which indicates that the encoder network module does not significantly affect the ArSSR model for the SISR tasks. In Table \ref{table-ex3}, we demonstrate the quantitative results of the ArSSR models with the three different encoder networks. In terms of PSNR and SSIM \cite{SSIM}, the ArSSR model with the ResCNN \cite{ResCNN} encoder network achieves the highest scores for all the scales. In terms of LPIPS \cite{LPIPS} and PSI \cite{PSI}, the ArSSR model with the RDN \cite{RDN} encoder network obtains the best performance for all the scales. While for LPC-SI \cite{LPC-SI}, the ArSSR model with the SRResNet \cite{SRGAN} encoder network produces the highest scores for all the scales. In summary, the three models achieve very close SR performance. These quantitative results further suggest that the ArSSR model is not significantly affected by the encoder network module.
\section{Conclusion}
\label{section-conclusion}
\par In this work, we proposed ArSSR model to conduct SISR tasks of arbitrary up-sampling scales for 3D MR images. Instead of learning the non-linear mapping from LR image to HR image as in conventional CNN-based SISR models \cite{SRCNN, DCSRN, ResCNN}, a continuous implicit voxel function in ArSSR model was used to represent the MR images of different spatial resolutions in the SISR tasks. Compared with the mapping from LR images to HR images, the implicit voxel function directly modeled the continuity of images in the spatial domain and thus recovers finer image details. Besides, by combining a convolution encoder network with the implicit voxel function, our ArSSR was also able to integrate the local semantic information in latent space as those in the CNN-based SISR models, which further improves the SR accuracy. The results from experiments on the two simulation datasets (a large public dataset with 1113 subjects and a small lesion brain dataset with 17 subjects) showed that for the SISR tasks of several different up-sampling scales, the single ArSSR model outperformed other popular CNNs-based methods in terms of both the HR fine details and image sharpness. It is worth noticing that the ArSSR model trained on the HCP-1200 dataset \cite{HCP-1200} of healthy subjects is also able to produce pleasing SR results on the lesion brain dataset, which means the proposed model archives a powerful generalization for different image patterns. Moreover, on the real collected dataset, the SISR results from the single ArSSR model provided finer anatomical details and sharper image contrast compared to the HR reference image with low SNR, which demonstrated that ArSSR model was capable of more broad real scene applications. In summary, the single ArSSR model based on the implicit neural representation can conduct HR MR image reconstruction at arbitrary scales, which would substantially improve the practicality of HR MRI acquisition.
\section{Limitation}
\par Although the proposed ArSSR model achieves superior performance for the arbitrary scale T1-weighted MRI SR tasks, there still exists a major limitation: The generalization ability of our ArSSR model is limited by the training data. It is known that the performance of the supervised deep learning SR models mostly depends on the scale and data distribution of HR-LR pairs in the training dataset (i.e., a large-scale training dataset that includes more types of MR images generally provides better performance). Therefore, we expect that the supervised ArSSR model trained on T1w brain MR images will suffer from a performance drop for other types of MR images (e.g., different acquisition sequences, organs, etc.). We hold that multi-model image data could improve the model generalization further. This is our future work.
\section{Acknowledgment}
\par The authors would like to thank Jiangfeng Hu and Fan Xu for careful and helpful proofreading.
\bibliographystyle{ieeetr}
\bibliography{ref}

\end{document}